\documentclass[a4paper,10pt,twocolumn]{elsarticle}

\usepackage{amsmath}
\usepackage{amssymb}
\usepackage{graphicx}
\usepackage{subcaption}
\usepackage{url}
\usepackage[top=3em, bottom=5em, left=3em, right=3em]{geometry}
\usepackage{tabularx}
\usepackage{supertabular}
\usepackage{multirow}
\usepackage{boxedminipage}
\usepackage{eurosym}

\newtheorem{remark}{Remark}

\begin{document}

\onecolumn

\begin{frontmatter}

\title{A Community Microgrid Architecture with an Internal Local Market}

\author[montefiore]{Bertrand~Corn\'elusse\corref{cor1}}
\ead{bertrand.cornelusse@uliege.be}

\author[siena]{Iacopo~Savelli}
\ead{savelli@diism.unisi.it}

\author[siena]{Simone~Paoletti}
\ead{paoletti@diism.unisi.it}

\author[siena]{Antonio~Giannitrapani}
\ead{giannitrapani@diism.unisi.it}

\author[siena]{Antonio~Vicino}
\ead{vicino@diism.unisi.it}

\cortext[cor1]{Corresponding author}

\address[montefiore]{Department of Electrical Engineering and Computer Science,
University of Li\`ege, Belgium}

\address[siena]{Dipartimento di Ingegneria dell'Informazione e Scienze Matematiche,
Universit\`a di Siena, Italy}

\begin{abstract}
This work fits in the context of community microgrids, where members of a community can exchange energy and services among themselves, without going through the usual channels of the public electricity grid. We introduce and analyze a framework to operate a community microgrid, and to share the resulting revenues and costs among its members. A market-oriented pricing of energy exchanges within the community is obtained by implementing an internal local market based on the marginal pricing scheme. The market aims at maximizing the social welfare of the community, thanks to the more efficient allocation of resources, the reduction of the peak power to be paid, and the increased amount of reserve, achieved at an aggregate level. A community microgrid operator, acting as a benevolent planner, redistributes revenues and costs among the members, in such a way that the solution achieved by each member within the community is not worse than the solution it would achieve by acting individually. In this way, each member is incentivized to participate in the community on a voluntary basis. The overall framework is formulated in the form of a bilevel model, where the lower level problem clears the market, while the upper level problem plays the role of the community microgrid operator. Numerical results obtained on a real test case implemented in Belgium show around 54\% cost savings on a yearly scale for the community, as compared to the case when its members act individually.
\end{abstract}


\end{frontmatter}

\twocolumn


\newcommand{\BSSs}{\ensuremath{\mathcal{D}^{sto}}}
\newcommand{\OPSOC}{\ensuremath{s}}
\newcommand{\maxcharge}{\ensuremath{\overline{S}}}
\newcommand{\mincharge}{\ensuremath{\underline{S}}}
\newcommand{\chargerate}{\ensuremath{\overline{P}}}
\newcommand{\dischargerate}{\ensuremath{\underline{P}}}
\newcommand{\retentionRate}{\ensuremath{\eta^{\text{retention}}}}
\newcommand{\chargeEfficiency}{\ensuremath{\eta^{\text{cha}}}}
\newcommand{\dischargeEfficiency}{\ensuremath{\eta^{\text{dis}}}}
\newcommand{\initialCharge}{\ensuremath{S^\text{init}}}
\newcommand{\minEndCharge}{\ensuremath{\underline{S}^\text{end}}}
\newcommand{\maxEndCharge}{\ensuremath{\overline{S}^\text{end}}}
\newcommand{\charge}{\ensuremath{a^\text{cha}}}
\newcommand{\discharge}{\ensuremath{a^\text{dis}}}
\newcommand{\finalCharge}{\ensuremath{S^{\text{end}}}}
\newcommand{\BSSsFee}{\ensuremath{\gamma^\text{sto}}}

\newcommand{\devices}[1]{\ensuremath{\mathcal{D}^\text{#1}}}
\newcommand{\sheddableDevices}{\ensuremath{\devices{she}}}
\newcommand{\nonflexibleDevices}{\ensuremath{\devices{nfl}}}
\newcommand{\steerableDevices}{\ensuremath{\devices{ste}}}
\newcommand{\curtableDevices}{\ensuremath{\devices{nst}}}
\newcommand{\nonsteerableDevices}{\ensuremath{\devices{nst}}}

\newcommand{\users}{\ensuremath{\mathcal{U}}}
\newcommand{\profit}[1]{\ensuremath{J^\text{#1}}}
\newcommand{\optprofit}[1]{\ensuremath{J^{*, \text{#1}}}}
\newcommand{\profitshare}[2]{\ensuremath{r^{#1}(J^{*, \text{MU}}, #2)}}

\newcommand{\maxExportToGrid}{\ensuremath{E}_{u,t}^{\text{cap}}}
\newcommand{\maxImportFromGrid}{\ensuremath{I}_{u,t}^{\text{cap}}}

\newcommand{\phiShedMax}{\ensuremath{\varphi^{she}_{d,t}}}
\newcommand{\phiSteerMax}{\ensuremath{\varphi^{ste}_{d,t} }}
\newcommand{\phiChargeMax}{\ensuremath{\varphi^{cha}_{d,t}}}
\newcommand{\phiDischargeMax}{\ensuremath{\varphi^{dis}_{d,t}}}
\newcommand{\phiSocMax}{\ensuremath{\varphi^{socUp}_{d,t}}}
\newcommand{\phiSocMaxT}{\ensuremath{\varphi^{socUp}_{d,T}}}
\newcommand{\phiSocMin}{\ensuremath{\varphi^{socLo}_{d,t}}}
\newcommand{\phiSocMinT}{\ensuremath{\varphi^{socLo}_{d,T}}}
\newcommand{\priceCom}{\ensuremath{\pi^{\text{com}}}}
\newcommand{\priceComUPrime}{\ensuremath{\pi^{com}_{u',t}}}
\newcommand{\FlowDual}{\ensuremath{\mu_{t}}}
\newcommand{\phiGridCapExport}{\ensuremath{\varphi^{eCap}_{u,t}}}
\newcommand{\phiGridCapImport}{\ensuremath{\varphi^{iCap}_{u,t}}}
\newcommand{\phiPeak}{\ensuremath{\varphi^{peak}_{t}}}
\newcommand{\phiPeakNoInfluenceOnDSO}{\ensuremath{\varphi^{peak}_{u,t}}}
\newcommand{\omegaStorageCharge}{\ensuremath{\omega^{cha}_{d,t}}}
\newcommand{\omegaStorageDischarge}{\ensuremath{\omega^{dis}_{d,t}}}
\newcommand{\kappaReserveBssSocInc}{\ensuremath{\kappa^{s+}_{d,t}}}
\newcommand{\kappaReserveBssSocIncT}{\ensuremath{\kappa^{s+}_{d,T}}}
\newcommand{\phiReserveBssMaxInc}{\ensuremath{\varphi^{s+}_{d,t}}}
\newcommand{\kappaReserveBssSocDec}{\ensuremath{\kappa^{s-}_{d,t}}}
\newcommand{\kappaReserveBssSocDecT}{\ensuremath{\kappa^{s-}_{d,T}}}
\newcommand{\phiReserveBssMaxDec}{\ensuremath{\varphi^{s-}_{d,t}}}
\newcommand{\sigmaSOC}{\ensuremath{\sigma_{d,t}}}
\newcommand{\rhoReserveSymInc}{\ensuremath{\rho^{inc}_{t}}}
\newcommand{\rhoReserveSymDec}{\ensuremath{\rho^{dec}_{t}}}

\newcommand{\operatorincome}{\ensuremath{J}^\text{operator}}
\newcommand{\OPFee}{\ensuremath{\gamma^\text{com}}}

\newcommand{\peak}{\ensuremath{\overline{p}}}

\newcommand{\OPprice}[1]{\ensuremath{\pi^{\text{#1}}}}
\newcommand{\sheddingPrice}{\OPprice{she}}
\newcommand{\steerablePrice}{\OPprice{ste}}
\newcommand{\reservePrice}{\OPprice{res}}
\newcommand{\gridBuyPrice}{\OPprice{igr}}
\newcommand{\gridSalePrice}{\OPprice{egr}}

\newcommand{\OPcost}[1]{\ensuremath{c^\text{#1}}}

\newcommand{\exportGrid}{\ensuremath{e}^\text{gri}}
\newcommand{\importGrid}{\ensuremath{i}^\text{gri}}
\newcommand{\OPexchangeOut}{\ensuremath{e}^\text{com}}
\newcommand{\OPexchangeIn}{\ensuremath{i}^\text{com}}
\newcommand{\OPprod}{\ensuremath{p^{\text{prod}}}}
\newcommand{\OPcons}{\ensuremath{p^{\text{cons}}}}

\newcommand{\OPreserve}[1]{\ensuremath{r^\text{#1}}}
\newcommand{\reserveBSSInc}{\ensuremath{r^{s+}_{d,t}}}
\newcommand{\reserveBSSDec}{\ensuremath{r^{s-}_{d,t}}}

\newcommand{\OPduration}{\ensuremath{\Delta_T}}
\newcommand{\OPperiods}{\ensuremath{\mathcal{T}}}

\newcommand{\steer}{\ensuremath{a^\text{ste}}}
\newcommand{\nonSteerable}{\ensuremath{P^\text{nst}}}
\newcommand{\steerable}{\ensuremath{P^\text{ste}}}

\newcommand{\shed} {\ensuremath{a^\text{she}}}
\newcommand{\flexCons}{\ensuremath{c^\text{flex}}}
\newcommand{\flexStart}{\ensuremath{y^\text{flex}}}
\newcommand{\flexStarted}{\ensuremath{z^\text{flex}}}
\newcommand{\exclusiveGroup}{\ensuremath{\mathcal{G}}}
\newcommand{\nonFlexible}{\ensuremath{C^\text{nfl}}}
\newcommand{\flexibleLoadDuration}{\ensuremath{\text{duration}^\text{fl}}}
\newcommand{\flexibleLoadProfile}{\ensuremath{\text{profile}^\text{flex}}}
\newcommand{\flexibleLoadStartTime}{\ensuremath{\text{start}^\text{flex}}}
\newcommand{\flexibleLoadEndTime}{\ensuremath{\text{end}^\text{flex}}}
\newcommand{\flexibleLoadAcceptanceRatio}{\ensuremath{\underline{a}^\text{flex}}}
\newcommand{\flexibleLoadMustRun}{\ensuremath{\text{run}^\text{flex}}}
\newcommand{\flexibleEnergy}{\ensuremath{\text{E}^\text{flex}}}
\newcommand{\sheddable}{\ensuremath{C^\text{she}}}
\newcommand{\maxSheddingTime}{\ensuremath{\overline{A}^\text{she}}}

\newcommand{\weight}[1]{\ensuremath{w^\text{#1}}}

\newcommand{\forallt}{\ensuremath{\forall t \in \OPperiods}}
\newcommand{\forallu}{\ensuremath{\forall u \in \users}}
\newcommand{\foralluPrime}{\ensuremath{\forall u' \in \users}}
\newcommand{\forallb}{\ensuremath{\forall d \in \BSSs}}
\newcommand{\forallDShed}{\ensuremath{\forall d \in \sheddableDevices}}
\newcommand{\forallDFlex}{\ensuremath{\forall d \in \devices{fl}}}
\newcommand{\forallDCurt}{\ensuremath{\forall d \in \devices{nst}}}
\newcommand{\forallDSteer}{\ensuremath{\forall d \in \devices{ste}}}

\section*{Nomenclature}

\subsection*{\textbf{Acronyms}}
\begin{tabularx}{\textwidth}{l X}
DG & Distributed generation. \\
DSO & Distribution system operator.
\end{tabularx}

\subsection*{\textbf{Sets and indices}}
\begin{supertabular}{l p{0.8\columnwidth}}
Name & Description \\
	\hline
	$d$ & Index of a device.\\	
	$\devices{nfl}_u$ & Set of non-flexible loads belonging to entity $u \in \users$.\\
	$\devices{nfl}$ &   Set of all non-flexible loads, where $\devices{nfl} = \cup_u \devices{nfl}_u$.\\
	$\devices{nst}_u$ & Set of non-steerable generators belonging to entity $u \in \users$.\\
	$\devices{nst}$ &   Set of all non-steerable generators, where $\devices{nst} = \cup_u \devices{nst}_u$.\\
	$\devices{she}_u$ & Set of sheddable loads belonging to entity $u \in \users$.\\
	$\devices{she}$ &   Set of all sheddable loads, where $\devices{she} = \cup_u \devices{she}_u$.\\
	$\devices{ste}_u$ & Set of steerable generators belonging to entity $u \in \users$.\\
	$\devices{ste}$ &   Set of all steerable generators, where $\devices{ste} = \cup_u \devices{ste}_u$.\\
	$\devices{sto}_u$ & Set of storage devices belonging to entity $u \in \users$.\\	
	$\devices{sto}$ &   Set of all storage devices, where $\devices{sto} = \cup_u \devices{sto}_u$.\\
	$t$ & Index of a time period. \\
	$T$ & Number of time periods. \\
	$\OPperiods$ & Set of time periods, with $\OPperiods = \{1, 2, ..., T\}$. \\
	$u$ & Index of an entity of the community microgrid.\\	
	$\users$ & Set of all entities of the community.\\
\end{supertabular}

\subsection*{\textbf{Parameters}}
\begin{supertabular}{l p{0.65\columnwidth} l}
Name & Description & Unit \\
\hline
$\nonFlexible_{d,t}$ & Non-flexible power consumption at time period $t$, with $d \in \devices{nfl}$. & kW \\
$\sheddable_{d,t}$ & Flexible power consumption at time period $t$, with $d \in \devices{she}$. & kW\\

$\maxExportToGrid$   & Maximum power export to the grid at time period $t$, with $u \in \users$.& kW \\
$\maxImportFromGrid$ & Maximum power import from the grid at time period $t$, with $u \in \users$.& kW \\

$\nonSteerable_{d,t}$ & Non-steerable power generation at time period $t$, with $d \in \devices{nst}$. & kW \\
$\steerable_{d,t}$ & Steerable power generation at time period $t$, with $d \in \devices{ste}$. & kW \\

$\chargerate_{d}$ & Maximum charging power of battery~$d$, with $d \in \devices{sto}$. & kW \\
$\dischargerate_{d}$ & Maximum discharging power of battery~$d$, with $d \in \devices{sto}$. & kW \\

$\maxcharge_{d}$ & Maximum capacity of battery~$d$, with $d \in \devices{sto}$.& kWh \\
$\mincharge_{d}$ & Minimum capacity of battery~$d$, with $d \in \devices{sto}$.& kWh \\

$\initialCharge_{d}$ & Initial state of charge of battery~$d$, with $d \in \devices{sto}$. & kWh \\
$\finalCharge_{d}$ & Final state of charge of battery~$d$, with $d \in \devices{sto}$. & kWh \\

$\profit{SU}_u$ & Optimal profit of entity $u \in \users$ acting individually. & \euro\\

$\BSSsFee_{d}$ & Unitary cost for usage of battery~$d$, with $d \in \devices{sto}$. & \euro/kWh\\
$\OPFee$ & Unitary fee of the community microgrid operator. & \euro/kWh\\
$\chargeEfficiency_{d}$ & Charging efficiency of battery $d$, with $d \in \devices{sto}$. & /\\
$\dischargeEfficiency_{d}$ & Discharging efficiency of battery $d$, with $d \in \devices{sto}$.& /\\
$\OPprice{peak}$ & Unitary cost (penalty) for peak power. & \euro/kW \\
$\reservePrice$ & Unitary revenue for providing reserve capacity. & \euro/kW \\
$\sheddingPrice_{d,t}$ & Unitary cost of load shedding at time period~$t$, with $d \in \devices{she}$. & \euro/kWh \\
$\steerablePrice_{d,t}$ & Unitary cost of generating energy at time period~$t$, with $d \in \devices{ste}$. & \euro/kWh \\
$\gridSalePrice_t$ & Unit price of energy exported to the grid at time period~$t$. & \euro/kWh \\
$\gridBuyPrice_t$ & Unit price of energy imported from the grid at time period~$t$. & \euro/kWh \\
$\OPduration$ & Duration of a time period.&  hrs  \\
\end{supertabular}
\\~\\

\subsection*{\textbf{Variables}}
\begin{supertabular}{l l p{0.55\columnwidth} l}
Name & Range & Description & Unit \\
\hline
$\alpha$ & $\mathbb{R}_+$ & Slack variable to be maximized. & \euro \\
$\shed_{d,t} $ & $ [0,1]$ & Fraction of flexible power consumption that is shed at time period~$t$, with $d \in \devices{she}$.& /\\
$\steer_{d,t} $ & $ [0,1]$ & Fraction of steerable power generation that is activated at time period~$t$, with $d \in \devices{ste}$. & /\\
$\charge_{d,t} $ & $ [0,1]$ & Fraction of the maximum charging power used for battery~$d$ at time period~$t$, with $d \in \devices{sto}$.  & / \\
$\discharge_{d,t} $ & $ [0, 1]$ & Fraction of the maximum discharging power used for battery~$d$ at time period~$t$, with $d \in \devices{sto}$.  & / \\
$\OPexchangeOut_{u,t}$ & $ \mathbb{R}_+$ & Energy exported to the community at time period~$t$ by entity $u \in \users$. & kWh\\
$\exportGrid_{u,t}$ & $ \mathbb{R}_+$ & Energy exported to the grid at time period~$t$ by entity $u \in \users$. & kWh\\
$\OPexchangeIn_{u,t}$ & $ \mathbb{R}_+$ & Energy imported from the community at time period~$t$ by entity $u \in \users$. & kWh\\
$\importGrid_{u,t}$ & $ \mathbb{R}_+$ & Energy imported from the grid at time period~$t$ by entity $u \in \users$. & kWh\\
$\peak$ & $ \mathbb{R}_+ $ & Peak power of the microgrid over the planning horizon~$\OPperiods$. & kW\\
$\peak_{u}$ & $ \mathbb{R}_+ $ & Contribution to the peak power of the microgrid assigned to entity $u \in \users$. & kW\\
$\OPreserve{sym}$ & $ \mathbb{R_+}$ & Symmetric power reserve provided by the microgrid through all periods of the planning horizon~$\OPperiods$. & kW\\
$\OPreserve{sym}_u$ & $ \mathbb{R_+}$ & Contribution to the symmetric power reserve of the microgrid assigned to entity $u \in \users$. & kW\\
$\OPreserve{inc}_{u,t}$ & $\mathbb{R_+}$ & Upward reserve of power available at time period~$t$ and provided by entity $u \in \users$.& kW\\
$\OPreserve{dec}_{u,t}$ & $\mathbb{R_+}$ & Downward reserve of power available at time period~$t$ and provided by entity $u \in \users$.& kW\\
$\reserveBSSInc$ & $\mathbb{R_+}$ & Upward reserve of power available at time period~$t$ and provided by storage device~$d \in \devices{sto}$. & kW \\
$\reserveBSSDec$ & $\mathbb{R_+}$ & Downward reserve of power available at time period~$t$ and provided by storage device~$d \in \devices{sto}$. & kW \\
$\OPSOC_{d,t}$ & $ [\mincharge_{d}, \maxcharge_d]$ & State of charge of battery~$d$ at time period~$t$, with $d \in \devices{sto}$. & kWh\\
$\profit{}_u$ & $\mathbb{R}$ & Total profit of entity $u \in \users$ acting in the community. & \euro\\
$\profit{energy}_u$ & $\mathbb{R}$ &  Energy component of the profit of entity~$u \in \users$. & \euro\\
$\profit{peak}_u$ & $\mathbb{R_-}$ & Peak power component of the profit of entity~$u \in \users$. & \euro\\
$\profit{reserve}_u$ & $\mathbb{R_+}$ & Reserve component of the profit of entity~$u \in \users$. & \euro\\
$\profit{*}$ & $\mathbb{R}$ & Optimal profit of the community microgrid. & \euro\\
\end{supertabular}

\subsection*{\textbf{Dual Variables}}
\begin{supertabular}{l l p{0.55\columnwidth}}
Name & Range & Corresponding constraint \\
\hline
$\kappaReserveBssSocInc$ & $\mathbb{R}_+$ & Storage upward reserve, state of charge.\\
$\kappaReserveBssSocDec$ & $\mathbb{R}_+$ & Storage downward reserve, state of charge.\\
$\FlowDual$ & $\mathbb{R}$ & Community energy balance.\\
$\priceCom_{u,t}$ & $\mathbb{R}$ & Entity energy balance.\\
$\rhoReserveSymInc$ & $\mathbb{R}_+$ & Symmetric reserve, upward reserve.\\
$\rhoReserveSymDec$ & $\mathbb{R}_+$ & Symmetric reserve, downward reserve. \\
$\sigmaSOC$ & $\mathbb{R}$ & Dynamics of battery state of charge.\\
$\phiShedMax$ & $ \mathbb{R}_+$ & Maximum load shedding.\\
$\phiSteerMax$ & $ \mathbb{R}_+$ & Maximum steerable generation.\\
$\phiChargeMax$ & $ \mathbb{R}_+$ & Maximum storage charging.\\
$\phiDischargeMax $ & $ \mathbb{R}_+$ &  Maximum storage discharging. \\
$\phiSocMax$ & $ \mathbb{R}_+$ & Maximum battery state of charge.\\
$\phiSocMin$ & $ \mathbb{R}_+$ & Minimum battery state of charge. \\
$\phiGridCapExport$ & $\mathbb{R}_+$ & Maximum energy export to the grid.\\
$\phiGridCapImport$ & $\mathbb{R}_+$ & Maximum energy import from the grid. \\
$\phiPeak$ & $\mathbb{R}_+$ & Community peak power.\\
$\phiReserveBssMaxInc$ & $\mathbb{R}_+$ & Storage upward reserve, power.\\
$\phiReserveBssMaxDec$ & $\mathbb{R}_+$ & Storage downward reserve, power.\\
$\zeta_d$ & $\mathbb{R}$ & Final battery state of charge.\\
\end{supertabular}

\section{Introduction}\label{sec:Introduction}
The increasing penetration of distributed generation (DG) from renewable energy sources and energy storage systems in distribution networks paves the way to new market models that favor a local usage of the generated electricity \cite{Morales2014}. In this context, microgrids are gaining increasing popularity as an architecture capable of making a more efficient use of resources at a local level \cite{Hatziargyriou2014}, and maximizing the local consumption of electricity generated in a distributed manner \cite{Hirsch2018}. When interconnected to the public grid, microgrids may also provide services, such as peak shaving and power balance.

\begin{figure}[!t]
	\centering
	\includegraphics[width=\linewidth]{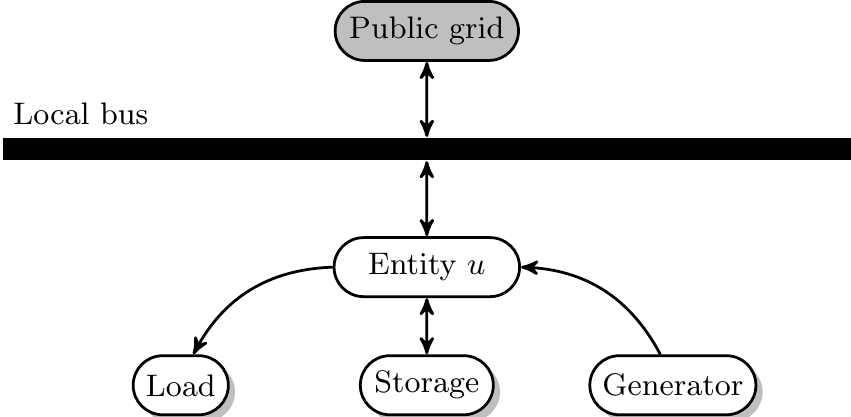}
	\caption{Entity of a community, device types, and energy flows.}
	\label{fig:SU}
\end{figure}

The contribution of this paper focuses on \emph{community microgrids}, where members of the community (termed \emph{entities} in the following) decide to pool their resources (generation, load and/or storage devices) to reduce their costs, increase their revenues, and achieve a more efficient use of their assets. A schematic representation of an entity is shown in Fig.~\ref{fig:SU}. The entities of the community are assumed to be connected to the same local bus, through which they exchange energy among themselves and with the public grid. After introducing a conceptual architecture of the community microgrid, this paper develops the model of an internal local market, based on the marginal pricing scheme, whose aim is to maximize the social welfare of the community.

\subsection{Related work}\label{sec:LiteratureReview}
Microgrid energy markets provide small-scale prosumers with a market platform to trade locally generated energy within their community. In some cases, the trading takes place without the need of central intermediaries. Blockchain-based local energy trading is proposed in \cite{brooklynMicrogrid2018Blockchain}, where prosumers can trade self-produced energy in a peer-to-peer fashion. A case study based on a real community microgrid project in Brooklyn is also reported. In \cite{khazaei2018indirect}, a non-cooperative game arises from the transferrable payoff allocation mechanism designed to aggregate renewable power producers in a two-settlement power market.

In most cases, the internal community market is managed by a third party. A coupled microgrid power and reserve capacity planning problem is considered in \cite{Quashie2018}. In the proposed bilevel formulation, the upper level problem represents a microgrid planner whose goal is to minimize its planning and operational cost, while the lower level problem represents a distribution system operator (DSO), whose primary duty is to ensure reliable power supply. In \cite{energySharingPeerToPeer2017Liu}, a peer-to-peer energy sharing microgrid model for photovoltaic prosumers is proposed. A virtual entity, named energy sharing provider, coordinates the sharing activities. An internal pricing system is designed, where the community prices are heuristically determined, based on the proportion between supply and demand in the energy sharing zone. Three potential market models for prosumers, i.e., prosumer-to-grid integration, peer-to-peer models, and community groups, are discussed in \cite{parag2016electricity}. Different decision making problems within a community of energy prosumers are considered in \cite{Olivier2017}, where a discussion on centralized and distributed control schemes is also provided. The concept of transactive electricity grid is analyzed in \cite{NIST2016}, where prosumers are actively engaged in the transaction of electrical energy, and new operators can be present to provide and manage innovative services.

Fairness is an important concept in microgrid energy markets, to sustain a long-lasting aggregation of different entities driven by self-interests. A definition of fairness borrowed from the field of coalitional game theory is considered in \cite{zhao2018solar}. A market with transferable payoffs is proposed, whose competitive equilibrium offers an \emph{in the core} allocation mechanism for the prosumers in an aggregated microgrid. A mathematical formulation for a fair benefit distribution among participants in a microgrid is presented in \cite{Zhang2013}, where a central operator decides the best solution by using a Nash bargaining model, assuming discrete price levels for market prices. In \cite{energyCollectives2018MoretPinson}, a distributed market structure is proposed, where all prosumers are in charge of optimizing their assets individually. Optimality is achieved as prosumers are coordinated by a non-profit virtual node, called community manager. Since the latter is also envisaged as a guarantor of the common goals within the community, assessment of fairness among market participants plays a fundamental role.

Other contributions in the literature on prosumer communities focus on specific aspects, such as the integration of energy storage, demand response and mixed energy types. The role of storage devices in peer-to-peer communities is addressed in \cite{luth2018local}, where the problems of determining the value of storage, and the most suitable market configurations, are considered. In particular, two market designs are analysed. In the first one, storage devices are located at the house level, while in the second, the storage is a centralized device at the community level. Similarly, the value of storage units in a real community energy system is investigated in \cite{granado2016synergy}, by analysing the conditions under which energy storage is valuable for a mix of local generation units supplying heating and electric loads. Techno-economic analyses of community energy storage for residential prosumers with smart appliances are proposed in \cite{technoEconomicAnalysisEnergyStorage2018Stelt} and \cite{communityEnergyStorageSmartChoide2018Barbour}, where economic indicators are evaluated under different scenarios. A community-based microgrid model, integrating wind turbines, photovoltaics and combined heat-and-power generation, is introduced in \cite{genericModelCommunityTurbinePV2013Ma}. Reference~\cite{hierarchicalManagementCommunity2015Xu} presents a hierarchical method for an integrated community energy system with demand response and combined heat-and-power units. The described approach is composed of a day-ahead scheduling system and two-layers for intra-hour adjustments, with different objectives representing the operating cost minimization and the tie-line power smoothing. Pricing of co-generated electricity and heat in local communities is also considered in \cite{Bohman1987}.

Notwithstanding, the problem of determining a market-oriented price for the trades within a community microgrid is still an open issue. This is the main thread motivating the contribution of this paper. Moreover, the paper tries to address existing gaps as concerns quantifying the monetary value of different services used by the community (e.g., a third-party operator managing the community market, and storage services).

\subsection{Paper contribution}\label{sec:PaperContribution}
The aim of this paper is to introduce a community microgrid conceptual architecture, with an internal market which shares the benefits of the community among the participating entities, by ensuring that none of them is penalized with respect to acting individually. The executed quantities and the market prices are determined by applying a social welfare maximization approach, within the marginal pricing framework, which ensures the efficient allocation of resources \cite{schweppe1988spot}\nocite{schweppeCaramanis1982optimal,mas1995microeconomic,rubinfeld2013microeconomics}--\cite{biggar2014}. By enforcing a set of non-discriminatory sharing policies, the revenues from the community reserve and the community peak power costs are shared among the entities. As a consequence, each entity can benefit from joining the community due to the following reasons:
\begin{itemize}
\item the more efficient allocation of resources, making it possible to trade energy at more favorable prices;
\item the provision of reserve on aggregate basis, and
\item the reduction of the peak power cost due to the netting effect of the other participants.
\end{itemize}

Summarizing, the main novelties of this paper in the context of community microgrids, are the following:
\begin{itemize}
\item the design of a local market for community microgrids, where the prices are determined through a social welfare maximization approach fulfilling the marginal pricing framework;
\item the introduction of a Pareto superior-type criterion to guarantee non-penalizing conditions for the entities in the community, thus stimulating participation on a voluntary basis;
\item the proposal of a benevolent planner operating the community, who fairly shares the profits of the community among the entities based on non-discriminatory rules;
\item the possibility to appraise different services used by the community, such as the storage and the community microgrid operator.
\end{itemize}

The remainder of the paper is organized as follows. Section~\ref{sec:comm_architecture} describes the considered community microgrid architecture. Section~\ref{sec:TheModel} formalizes the proposed community microgrid framework as a bilevel model, and discusses different strategies for its solution. Section~\ref{sec:NumericalResults} reports some toy examples to demonstrate the main features of the model. A test case inspired by the MeryGrid project \cite{cornelusse2017efficient}, a real pilot project currently under implementation in Belgium, is also presented. It shows the significant cost savings on a yearly scale for the community members, achieved by the proposed framework in a real world problem. Section~\ref{sec:Conclusions} summarizes the main findings, and highlights ideas for further work.

\section{The community microgrid architecture} \label{sec:comm_architecture}
\begin{figure}[tb]
\centering
\begin{subfigure}[b]{\linewidth}
    \centering
    {\includegraphics[width=\linewidth]{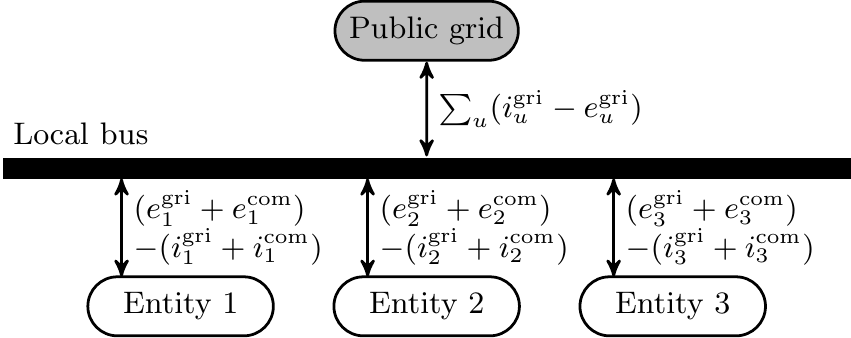}}
    \caption{Real energy flows}
    \label{fig:MU_real}
\end{subfigure}
\\[1em]%
\begin{subfigure}[b]{\linewidth}
    \centering
	{\includegraphics[width=\linewidth]{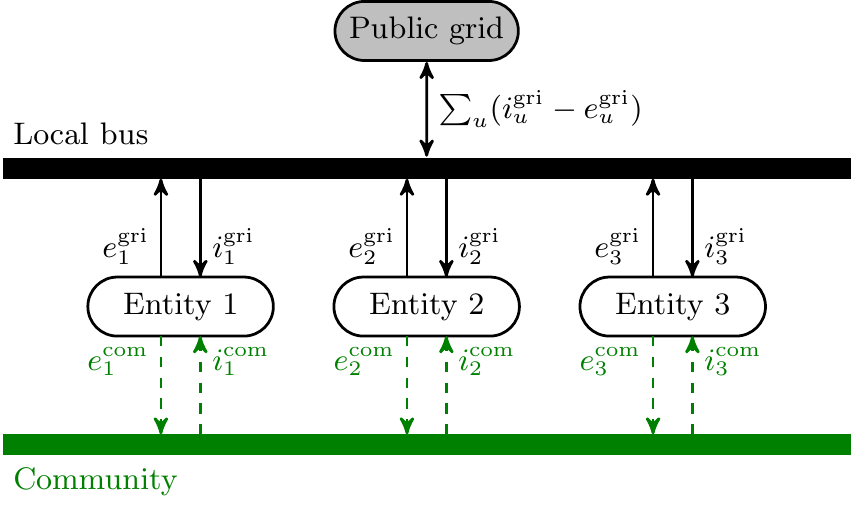}}
	\caption{Virtual energy flows}
	\label{fig:MU_virtual}
\end{subfigure}
\caption{Schematic representations of the entities and of the community.} \label{fig:com_schematic}
\end{figure}

A community microgrid is a collection of entities that exchange energy and services according to the rules of the community. Each entity is characterized by its own generation, load and/or storage devices, and is assumed to be connected to the public grid through a local bus, as shown in Fig.~\ref{fig:SU}. The local bus belongs to the public grid. When several entities are connected to the same local bus (see Fig.~\ref{fig:MU_real}), the community microgrid is a virtual layer handling the energy flows between entities, that do not cross the boundary of the local bus (see Fig.~\ref{fig:MU_virtual}). In this setting, $\OPexchangeOut_{u}$ and $\OPexchangeIn_{u}$ denote the energy exported to and imported from the community by entity~$u$, respectively\footnote{For the sake of simplicity, we omit for now the time index, see the Nomenclature.}. These energy flows represent the additional degrees of freedom offered by the community to the entities, enabling them to exchange energy among themselves. Letting $\exportGrid_{u}$ and $\importGrid_{u}$ be the energy exported to and imported from the grid by entity~$u$, respectively, the net energy flowing from entity~$u$ to the local bus amounts to $(\exportGrid_{u}+\OPexchangeOut_{u})-(\importGrid_{u}+\OPexchangeIn_{u})$. Since the net energy flowing from the grid to the local bus amounts to $\sum_{u\in\users} (\importGrid_{u}-\exportGrid_{u})$, where $\users$ is the set of all the entities, it must hold that $\sum_{u\in\users} (\OPexchangeIn_{u}-\OPexchangeOut_{u}) = 0$, corresponding to the energy balance at the community level. Figure~\ref{fig:MU_real} shows the actual energy flows in the considered distribution system. It can be also useful to visualize the virtual energy flows as shown in Fig.~\ref{fig:MU_virtual}, where the community is represented as a fictitious layer connecting the entities. %

Joining a community microgrid may bring several advantages to its participants:
\begin{itemize}
\item energy exchanges within the community can be executed at a price that is more advantageous than the external grid price;
\item reserve can be pooled over the entities, and exchanged with the grid at the community level;
\item the peak power penalty can be reduced, since the peak to be penalized is the peak of the community, rather than the sum of the peaks of the single entities, taken individually.
\end{itemize}
To achieve the aforementioned benefits, this paper focuses on the operational planning stage that optimizes day-ahead decisions with time periods of $\OPduration$~hours, given the grid energy prices, and the consumption and generation forecasts of the entities. An operator manages the community to maximize its social welfare, by optimizing the energy flows and the interactions among the entities and with the public grid. The operator must be contractually bound to the DSO, and its perimeter must be clearly defined. It must also be trusted by, and be contractually bound to, the community entities. The latter pay a fee to the operator for the remuneration of its activity.

It is stressed that, when entities aggregate in a community, the peak power penalty, which remunerates the DSO for the capacity of the public grid, is applied to the aggregate net energy flow $\sum_{u\in\users} (\importGrid_{u}-\exportGrid_{u})$. In this way, a single entity could see a decrease of its peak power costs, even if it does not lower its individual peak. However, adopting a peak penalty scheme of this type at the community level, would likely force the DSO to increase the peak penalty tariff. This, in turn, should incentivize community members to decrease and to desynchronize their peak power consumptions, which translates into lesser reinforcement costs for the DSO. Another positive impact of the community microgrids for the DSOs is that the coordinated optimization and control of several entities offer a simpler interface, e.g., for the procurement of flexibility services.

For the energy exchanged with the external grid, entities are subject to the same mechanism as if they would not be part of a community (electricity tariffs, fees, taxes, etc.). In practice, this requires a correction of the data communicated to operators and retailers for invoicing purposes. Indeed, assuming two meters per entity, the one will measure the energy flowing from the entity to the local bus, namely $\exportGrid_{u}+\OPexchangeOut_{u}$, the other will measure the energy flowing from the local bus to the entity, namely $\importGrid_{u}+\OPexchangeIn_{u}$. Both measurements should be corrected with the information provided by the community microgrid operator (in particular, with the internal energy flows $\OPexchangeOut_{u}$ and $\OPexchangeIn_{u}$) to obtain the correct values $\exportGrid_{u}$ and $\importGrid_{u}$ to be invoiced to entity~$u$.

\section{The optimization model}\label{sec:TheModel}
This section describes the optimization model designed to solve the community microgrid market clearing problem, and to share the corresponding benefits among the entities of the community. The problem is formulated in Section~\ref{sec:BilevelProgramming} as a nonlinear bilevel model. Practical aspects on how to tackle the solution of the proposed bilevel model are discussed in Section~\ref{sec:SingleLevel}.

\subsection{The bilevel model}\label{sec:BilevelProgramming}
A bilevel model is a mathematical program composed of two nested optimization problems, termed upper and lower level \cite{bard1998practical}. Formally,
\begin{align}
&\max_{x \in \mathcal{X}} F(x, y^*)\label{bilevel_sketch_1}\\
&\text{s.t.}\quad y^* \in \arg \max_{y \in \mathcal{Y}} f(y;x) \, , \label{bilevel_sketch_2}
\end{align}
where $F$ and $f$ are the objective functions of the upper and lower level problems \eqref{bilevel_sketch_1} and \eqref{bilevel_sketch_2}, respectively. In general, the optimizer $y^*$ and the feasible set $\mathcal{Y}$ of the lower level depend on the unknown $x$ of the upper level. In turn, the feasible set $\mathcal{X}$ of the upper level may depend on $y^*$.

Historically, bilevel programming was used in the field of game theory to cast Stackelberg games \cite{Sinha2018}. These problems represent a class of non-cooperative leader-follower games, where the upper and lower level objective functions are different, and represent conflicting interests. Conversely, in power system economics \cite{conejo2012complementarity,blanco2014consumer,Feijoo2014,iacopoAPEN2018}, bilevel programming is typically used to access dual variables, which represent market prices within the marginal pricing framework \cite{schweppe1988spot}. This is the main reason to adopt a bilevel formulation in this paper. In the proposed bilevel model, the lower level solves the community microgrid market clearing problem, by determining:
\begin{itemize}
\item the executed quantities,
\item the community prices,
\item the community symmetric reserve, and
\item the community peak power,
\end{itemize}
with the aim of maximizing the social welfare of the community. Then, the role of the upper level is to share among the entities the profits of the community, by ensuring that no entity is penalized with respect to acting individually. Indeed, a community is composed of several entities driven by self-interests. In order to achieve a long-lasting community microgrid, each entity should be stimulated to participate in the community on a voluntary basis. This is achieved by enforcing the condition
\begin{equation} \label{eq:ParetoCondition}
\profit{}_u \geq \profit{SU}_u \qquad \forallu,
\end{equation}
where $\profit{SU}_{u}$ is the optimal profit of entity~$u$ when acting individually, while $\profit{}_{u}$ is the profit of entity~$u$ within the community. As a consequence, the upper level can be seen as a \textit{benevolent planner}, that manages the community microgrid in the best interest of all the participants. If at least one inequality in \eqref{eq:ParetoCondition} is satisfied strictly, the state of the community is termed \textit{Pareto superior} to the state when entities act individually \cite{mas1995microeconomic}.

\subsubsection{Lower level problem}\label{sec:LowerLevel}
This section describes the lower level problem \eqref{bilevel_sketch_2} of the proposed bilevel model. The vector $y$ of decision variables is composed of the community symmetric reserve $\OPreserve{\text{sym}}$, the community peak power $\peak$, the executed quantities $\exportGrid_{u,t}$, $\importGrid_{u,t}$, $\OPexchangeOut_{u,t}$, $\OPexchangeIn_{u,t}$, the community prices $\priceCom_{u,t}$, the upward and downward power reserves $\reserveBSSInc$, $\reserveBSSDec$, the fractions $\shed_{d,t}$, $\steer_{d,t}$ of shed demand and steered generation, the fractions of charging and discharging power $\charge_{d,t}$, $\discharge_{d,t}$, and the storage states of charge $\OPSOC_{d,t}$.

The objective function $f$ is defined as follows:
\begin{align}
& -\sum_{u \in \users}\sum_{t \in \OPperiods} \bigg(\sum_{d\in \sheddableDevices_u} \sheddingPrice_{d,t} \sheddable_{d,t} \OPduration \shed_{d,t} + \sum_{d\in \steerableDevices_u} \steerablePrice_{d,t}  \steerable_{d,t} \OPduration \steer_{d,t}\notag\\
&\hspace{5em} -\gridSalePrice_t \exportGrid_{u,t}+\gridBuyPrice_t \importGrid_{u,t} + \OPFee \left( \OPexchangeOut_{u,t} + \OPexchangeIn_{u,t} \right) \notag\\
&\hspace{5em} + \sum_{d\in\BSSs_{u}}
\BSSsFee_{d} \OPduration \left(\chargerate_{d} \chargeEfficiency_{d} \charge_{d,t} + \frac{\dischargerate_{d}}{\dischargeEfficiency_d} \discharge_{d,t} \right) \bigg) \notag\\
&+\reservePrice \OPreserve{sym} \notag \\
&- \OPprice{peak} \peak. \label{lower:objective_function}
\end{align}
It represents the social welfare of the community, composed of three terms. The first term is the summation $-\sum_{u \in \users}(\ldots)$, which takes into account different revenues and costs related to energy flows\footnote{In \eqref{lower:objective_function}, revenues are positive quantities, while costs are negative quantities.}: the costs of shed demand and steered generation, the revenues from selling energy to the grid, the costs of purchasing energy from the grid, the fees paid to the community microgrid operator, and the costs for using storage. The second term is $\reservePrice \OPreserve{sym}$, representing the revenue collected by the community for providing reserve to the grid. Finally, the third term is the cost $-\OPprice{peak}\peak$, paid for the  community peak power $\peak$ over the time horizon $\OPperiods$.

The feasible set $\mathcal{Y}$ of the lower level problem \eqref{bilevel_sketch_2} is defined by different sets of constraints. In the following, a variable between square brackets represents the dual variable of the corresponding constraint. The first set of constraints bounds some of the lower level decision variables:
\begin{align}
&\steer_{d,t} \leq 1     &&\forallDSteer,\forallt     &&[\phiSteerMax \geq 0] \label{lower:steerable_upper_bound}\\
&\shed_{d,t} \leq 1      &&\forallDShed,\forallt     &&[\phiShedMax \geq 0]\\
&\charge_{d,t} \leq 1    &&\forallb,\forallt     &&[\phiChargeMax \geq 0]  \\
&\discharge_{d,t} \leq 1 &&\forallb,\forallt     &&[\phiDischargeMax \geq 0] \\
&\OPSOC_{d,t} \leq \maxcharge_d &&\forallb, \forallt &&[\phiSocMax \geq 0] \label{lower:soc_upper_bound} \\
&-\OPSOC_{d,t} \leq -\mincharge_d &&\forallb, \forallt. &&[\phiSocMin \geq 0] \label{lower:soc_lower_bound}
\end{align}
In particular, constraints~\eqref{lower:soc_upper_bound} and \eqref{lower:soc_lower_bound} impose that the state of charge of each storage unit cannot exceed its upper and lower bounds $\maxcharge_d$ and $\mincharge_d$.

The second set of constraints is related to energy flows inside and outside the community.
\begin{align}
&\exportGrid_{u,t} - \importGrid_{u,t} + \OPexchangeOut_{u,t} - \OPexchangeIn_{u,t}  \notag\\
&- \OPduration \bigg(\sum_{d\in \nonsteerableDevices_u} \nonSteerable_{d,t} + \sum_{d\in \steerableDevices_u} \steer_{d,t} \steerable_{d,t}\bigg) \notag\\
& + \OPduration \bigg(\sum_{d\in\nonflexibleDevices_u} \nonFlexible_{d,t}
+ \sum_{d\in \sheddableDevices_u}(1 - \shed_{d,t}) \sheddable_{d,t} \bigg) \notag\\
&+ \OPduration \sum_{d\in\BSSs_u} \left(\chargerate_d \charge_{d,t} - \dischargerate_d \discharge_{d,t} \right)  = 0 \notag\\
& \hspace{8em} \forallu, \forallt \qquad [\priceCom_{u,t} \in \mathbb{R}] \label{lower:power_balance}\\
&\sum_{u \in \users} \left( \OPexchangeIn_{u,t} - \OPexchangeOut_{u,t} \right)  = 0  \qquad \forallt \qquad [\FlowDual \in \mathbb{R}] \label{lower:export_import_balance} \\
&(\exportGrid_{u,t} - \importGrid_{u,t})/\OPduration \leq \maxExportToGrid   \quad \forallu, \forallt \quad [\phiGridCapExport \geq 0] \label{lower:export_cap}\\
&(\importGrid_{u,t} - \exportGrid_{u,t})/\OPduration \leq \maxImportFromGrid \quad \forallu, \forallt \quad [\phiGridCapImport \geq 0] \label{lower:import_cap} \\
& \sum_{u \in \users} \left(\importGrid_{u,t} - \exportGrid_{u,t} \right) / \OPduration \leq \peak \quad \forallt. \qquad [\phiPeak \geq 0] \label{lower:peak_def}
\end{align}
Constraint~\eqref{lower:power_balance} defines the energy balance for each entity. The positive terms in the left-hand side of the equation represent the energy exported by the entity to the grid and the rest of the community, as well as the energy consumption of the loads connected to the entity, including the charging of storage units. On the other hand, the negative terms represent the energy imported by the entity from the grid and the rest of the community, as well as the energy provided by the generators connected to the entity, including the discharging of storage units. It is stressed that the dual variable $\priceCom_{u,t}$ of constraint \eqref{lower:power_balance} has an important economic interpretation within the marginal pricing framework \cite{schweppe1988spot,schweppeCaramanis1982optimal}. Indeed, being $\priceCom_{u,t}$ the dual variable of the energy balance constraint for entity~$u$, in the proposed formulation it represents the market price at which entity $u$ exchanges energy with the community at time $t$ \cite{litvinov2010locationalmarginalprices}. Constraint~\eqref{lower:export_import_balance} imposes the balance of the energy flows within the community at each time period. Constraints~\eqref{lower:export_cap} and \eqref{lower:import_cap} set limits to the net energy exported to and imported from the grid by each entity at each time period. Finally, constraint~\eqref{lower:peak_def} determines the community peak power $\peak$. The community peak power $\peak$ is defined as the maximum of the net power imported from the grid by the community over the time horizon $\OPperiods$. In this respect, the simultaneous import and export of different entities may have the effect of making the community peak power lower than the sum of the peaks of the entities when acting individually.

\begin{remark}
{\rm At the optimum of the lower level problem, $\exportGrid_{u,t}$ and $\importGrid_{u,t}$, as well as $\OPexchangeOut_{u,t}$ and $\OPexchangeIn_{u,t}$, cannot be simultaneously greater than 0. In other words, no simultaneous export to and import from the grid can occur for entity~$u$ over a given time period. The same holds for the energy exported to and imported from the community by entity~$u$. Both results are shown in \ref{appendix:proof_simultanelus_import_export}. \hfill$\Box$}
\end{remark}

The following set of constraints describes the dynamics of the state of charge for each storage unit:
\begin{align}
&s_{d,1} - \OPduration \left(\chargerate_{d} \chargeEfficiency_{d} \charge_{d,1} - \frac{\dischargerate_{d}}{\dischargeEfficiency_d} \discharge_{d,1} \right) = \initialCharge_{d} \notag\\
& \hspace{12em} \forallb  \qquad [\sigma_{d,1} \in \mathbb{R}] \label{lower:soc_1}\\
&s_{d,t} - s_{d, t-1} - \OPduration \left(\chargerate_{d} \chargeEfficiency_{d} \charge_{d,t} - \frac{\dischargerate_{d}}{\dischargeEfficiency_d} \discharge_{d,t} \right)=0 \notag&&&\\
&\hspace{4em} \forallb, \quad t \in \{2,\ldots,T\} \qquad [\sigmaSOC \in \mathbb{R}] \label{lower:soc_2} \\
&s_{d,T} = \finalCharge_{d} \qquad \forallb \qquad  [\zeta_d \in \mathbb{R}] \label{lower:soc_3}
\end{align}
In \eqref{lower:soc_1} and \eqref{lower:soc_3}, $\initialCharge_{d}$ and $\finalCharge_{d}$ are given parameters, representing the initial and final state of charge of storage unit~$d$, respectively. Notice that the dynamics in \eqref{lower:soc_1} and \eqref{lower:soc_2} take into account the charging and discharging efficiencies $\chargeEfficiency_{d}$ and $\dischargeEfficiency_d$ of the storage units.

The last set of constraints defines the community symmetric reserve $\OPreserve{sym}$:
\begin{align}
&\reserveBSSInc \leq \dfrac{\left( \OPSOC_{d,t} - \mincharge_{d}\right)\dischargeEfficiency_d}{\OPduration} \quad  \forallb,\forallt \quad [\kappaReserveBssSocInc \geq 0] \label{lower:reseve_BSS_inc_max_soc}\\
&\reserveBSSInc \leq \dischargerate_{d} (1-\discharge_{d,t}) \quad  \forallb,\forallt \quad   [ \phiReserveBssMaxInc\geq 0] \label{lower:reseve_BSS_inc_max_power}\\
&\reserveBSSDec \leq \dfrac{\left( \maxcharge_{d} - \OPSOC_{d,t} \right)}{\chargeEfficiency_d\OPduration} \quad \forallb,\forallt \quad  [\kappaReserveBssSocDec \geq 0] \label{lower:reseve_BSS_dec_max_soc}\\
&\reserveBSSDec \leq \chargerate_{d}(1-\charge_{d,t}) \quad \forallb,\forallt \quad [ \phiReserveBssMaxDec\geq 0] \label{lower:reseve_BSS_dec_max_power} \\
&\OPreserve{sym} \leq \sum_{u \in \users} \bigg(\sum_{d\in\BSSs_u} \reserveBSSInc + \sum_{d\in \steerableDevices_u}\steerable_{d,t} (1 - \steer_{d,t})\notag\\
&\hspace{4em} +\sum_{d\in \sheddableDevices_u} \sheddable_{d,t} (1-\shed_{d,t})\bigg) \qquad \forallt \qquad [\rhoReserveSymInc \geq 0] \label{lower:reserve_symmetric_1}\\
&\OPreserve{sym} \leq \sum_{u \in \users} \bigg(\sum_{d\in\BSSs_u} \reserveBSSDec + \sum_{d\in \steerableDevices_u} \steerable_{d,t} \steer_{d,t} \notag\\
&\hspace{4em} +\sum_{d\in \sheddableDevices_u} \sheddable_{d,t} \shed_{d,t} \bigg)\qquad \forallt \qquad [\rhoReserveSymDec \geq 0] \label{lower:reserve_symmetric_2}
\end{align}
The upward reserve $\reserveBSSInc$ provided by storage unit~$d$ at time~$t$ is bounded by both its current state of charge $\OPSOC_{d,t}$ through \eqref{lower:reseve_BSS_inc_max_soc}, and the remaining available discharging power through \eqref{lower:reseve_BSS_inc_max_power}. Inequalities \eqref{lower:reseve_BSS_dec_max_soc} and \eqref{lower:reseve_BSS_dec_max_power} enforce similar constraints to the downward reserve provided by each storage unit. Finally, constraints \eqref{lower:reserve_symmetric_1} and \eqref{lower:reserve_symmetric_2} define the community symmetric reserve as the minimum symmetric reserve available from the community over the time horizon $\OPperiods$. Notice that the reserve is provided not only by storage units, but also by steerable generators and sheddable loads. Other reserve schemes can be modeled, as long as the corresponding constraints remain linear in the decision variables of the lower level problem.

Summarizing, the lower level problem solves the community microgrid market clearing problem by maximizing the objective function~\eqref{lower:objective_function}, subject to the constraints \eqref{lower:steerable_upper_bound}-\eqref{lower:reserve_symmetric_2}. Notice that this is a linear program. By applying standard tools of duality theory in linear programming, the constraints and the objective function of the corresponding dual problem are derived in \ref{appendix:dual_contraints} and \ref{appendix:strong_duality}. Notice that solving the dual problem gives access to the community prices $\priceCom_{u,t}$, that are a key outcome of the market clearing process. In the following, the value of the objective function~\eqref{lower:objective_function} at the optimum of the lower level problem will be denoted by $\profit{*}$.

An important result that can be obtained from duality relations, is the following identity, which holds at the optimum of the lower level problem (see \ref{appendix:proof_sum_bilinear_terms} for the proof):~
\begin{equation} \label{lower:cost_identity}
\OPFee \sum_{u \in \users} \sum_{t \in \OPperiods} \left( \OPexchangeOut_{u,t} + \OPexchangeIn_{u,t} \right) = -\sum_{u \in \users} \sum_{t \in \OPperiods} \OPprice{com}_{u,t} \left( \OPexchangeOut_{u,t} - \OPexchangeIn_{u,t} \right).
\end{equation}
Intuitively, this can be explained because the monetary flows within the community offset each other, except for the part collected by the community operator, represented by its fee.

\subsubsection{Upper level problem}\label{sec:UpperLevel}
The role of the upper level is to share among the entities the optimal profit $\profit{*}$ of the community, while ensuring that no entity is penalized with respect to acting individually. To do this, we let
\begin{equation} \label{upper:total_profit_u}
\profit{}_u = \profit{energy}_u + \profit{reserve}_u + \profit{peak}_u
\end{equation}
be the total profit of entity~$u$ within the considered community microgrid framework. In \eqref{upper:total_profit_u}, the quantity $\profit{energy}_{u}$ takes into account the revenues and costs for entity~$u$ related to energy flows:
\begin{align}
&\hspace*{-0.8em}\profit{energy}_{u} = \notag\\
&=-\sum_{t \in \OPperiods} \bigg(\sum_{d\in \sheddableDevices_u}\sheddingPrice_{d,t} \sheddable_{d,t} \OPduration \shed_{d,t} + \sum_{d\in \steerableDevices_u}\steerablePrice_{d,t} \steerable_{d,t} \OPduration \steer_{d,t} \notag\\
&\hspace{4em} - \gridSalePrice_t \exportGrid_{u,t} + \gridBuyPrice_t \importGrid_{u,t} - \OPprice{com}_{u,t} \left( \OPexchangeOut_{u,t} - \OPexchangeIn_{u,t} \right) \notag\\
&\hspace{4em} + \sum_{d\in\BSSs_{u}}
\BSSsFee_{d} \OPduration \left(\chargerate_{d} \chargeEfficiency_{d} \charge_{d,t} + \frac{\dischargerate_{d}}{\dischargeEfficiency_d} \discharge_{d,t} \right) \bigg). \label{upper:J_energy_component}
\end{align}
Notice that the energy exchanges with the community, $\OPexchangeOut_{u,t}$ and $\OPexchangeIn_{u,t}$, are valued at the price $\OPprice{com}_{u,t}$, i.e., the market-clearing price for entity~$u$ at time~$t$. Moreover, in \eqref{upper:total_profit_u}, $\profit{reserve}_u$ represents the revenue ascribed to entity~$u$ for its contribution to reserve, and $\profit{peak}_u$ is the portion assigned to entity~$u$ of the cost paid by the community for the peak power. These quantities are defined as follows:
\begin{align}
\profit{reserve}_u &= \reservePrice \OPreserve{sym}_{u} \label{upper:J_reserve_profit} \\
\profit{peak}_u & = -\OPprice{peak} \peak_u, \label{upper:J_peak_cost}
\end{align}
where $\OPreserve{sym}_u \geq 0$ and $\peak_u \geq 0$ are the contributions to community reserve~$\OPreserve{sym}$ and peak power $\peak$ assigned to entity~$u$, satisfying the constraints:
\begin{align}
\OPreserve{sym} &= \sum_{u \in \users} \OPreserve{sym}_u \label{upper:reserve_symmetric_def} \\
\peak &= \sum_{u \in \users} \peak_u.  \label{upper:peak_sharing}
\end{align}
By summing \eqref{upper:total_profit_u} over all entities $u$, and exploiting \eqref{lower:cost_identity}, \eqref{upper:reserve_symmetric_def} and \eqref{upper:peak_sharing}, it is straightforward to obtain the identity:
\begin{equation} \label{upper:profit_comm}
\profit{*} = \sum_{u \in \users} \profit{}_u,
\end{equation}
which shows that the proposed framework totally shares the optimal profit $\profit{*}$ of the community among the entities.

In order to ensure that no entity is penalized with respect to acting individually, for each entity~$u$ the quantity $\profit{}_u$ in \eqref{upper:total_profit_u} has to be compared with the value $\profit{SU}_u$, representing the maximum profit that the entity would achieve over the time horizon $\OPperiods$ without joining the community. This value is computed for each entity by solving an optimization problem derived from the lower level problem of the community. Specifically, in \eqref{lower:objective_function}-\eqref{lower:reserve_symmetric_2}, all summations with respect to $u\in\users$ are removed, the index $u$ is fixed and refers to the entity considered, the energy exchanges $\OPexchangeOut_{u,t}$ and $\OPexchangeIn_{u,t}$ with the community are set to zero, and $\OPreserve{sym}$ and $\peak$ are replaced with $\OPreserve{sym}_u$ and $\peak_u$. Given the lower bounds $\profit{SU}_u$ for all entities, the requirement that all entities should benefit from joining the community, is translated into the following condition:
\begin{equation}\label{upper:pareto_superior_condition_alpha}
\profit{}_u \geq \profit{SU}_u + \alpha, \quad \forallu,
\end{equation}
where $\alpha \geq 0$ is a slack variable to be maximized. Notice that maximizing $\alpha$ corresponds to maximize $\min_{u} \big(\profit{}_u - \profit{SU}_u\big)$, i.e., the minimum profit improvement achieved by all the entities $u$ of the community. Since $\alpha \geq 0$, condition~\eqref{upper:pareto_superior_condition_alpha} generalizes condition~\eqref{eq:ParetoCondition}.

In order to satisfy \eqref{upper:pareto_superior_condition_alpha}, the upper level problem may act on the terms $\profit{reserve}_u$ and $\profit{peak}_u$ by deciding the quantities $\OPreserve{sym}_u$ and $\peak_u$, subject to \eqref{upper:reserve_symmetric_def} and \eqref{upper:peak_sharing}. Conversely, the term $\profit{energy}_u$ of \eqref{upper:total_profit_u} is fixed by the considered solution of the lower level problem. Notice, however, that the lower level problem may have multiple solutions, and the upper level problem explores all of them.

Additional constraints are introduced to define, together with \eqref{upper:reserve_symmetric_def}, the reserve sharing policy enforced by the community microgrid operator:
\begin{align}
&\OPreserve{inc}_{u,t} = \sum_{d\in\BSSs_u} \reserveBSSInc + \sum_{d\in \steerableDevices_u}\steerable_{d,t} (1 - \steer_{d,t})\notag\\
&\hspace{3em} + \sum_{d\in \sheddableDevices_u} \sheddable_{d,t} (1-\shed_{d,t}) \quad \forallu,\forallt \label{upper:reserve_inc} \\
&\OPreserve{dec}_{u,t}  = \sum_{d\in\BSSs_u} \reserveBSSDec  + \sum_{d\in \steerableDevices_u} \steerable_{d,t} \steer_{d,t} \notag\\
&\hspace{3em}+ \sum_{d\in \sheddableDevices_u} \sheddable_{d,t} \shed_{d,t}  \quad  \forallu,\forallt \label{upper:reserve_dec}\\
&\OPreserve{sym}_u  \leq \frac{1}{2} \left(\OPreserve{dec}_{u,t}+ \OPreserve{inc}_{u,t}\right) \quad \forallu, \forallt. \label{upper:reserve_one_half_rule}
\end{align}
Constraints \eqref{upper:reserve_inc} and \eqref{upper:reserve_dec} define the actual amount of upward and downward reserve provided by each entity $u$ at time period $t$. From \eqref{lower:reserve_symmetric_1} and \eqref{lower:reserve_symmetric_2}, it holds that $\OPreserve{sym} \leq \sum_{u \in \users} \OPreserve{inc}_{u,t}$ and $\OPreserve{sym} \leq \sum_{u \in \users} \OPreserve{dec}_{u,t}$ for all $t \in \OPperiods$. Constraint~\eqref{upper:reserve_one_half_rule} imposes that no entity is accounted for more than its average reserve contribution. This represents a possible non-discriminatory sharing policy rule, though more involved rules could be defined.

Summarizing, in the proposed formulation, the upper level problem \eqref{bilevel_sketch_1} is an optimization problem in the decision variables $\alpha$, $\OPreserve{inc}_{u,t}$, $\OPreserve{dec}_{u,t}$, $\OPreserve{sym}_u$ and $\peak_u$, with feasible set $\mathcal{X}$ defined by the constraints~\eqref{upper:reserve_symmetric_def}-\eqref{upper:reserve_one_half_rule}, and $\alpha \geq 0$. The objective function $F$ of the upper level coincides with the slack variable $\alpha$.

\subsection{Solution strategies}\label{sec:SingleLevel}
In the proposed bilevel formulation, the lower level problem is a linear program which does not depend on the decision variables of the upper level problem. Moreover, for a fixed solution of the lower level problem, the upper level problem is also a linear program. This implies that, if the solution of the lower level problem is unique, the bilevel model can be solved very efficiently as the cascade of two linear programs, one corresponding to the lower level, and the other corresponding to the upper level. Notice that uniqueness of the solution of the lower level problem can be checked a priori via standard tools in linear programming \cite{Mangasarian1979}.

If the lower level solution is not unique, the bilevel problem can be tackled by recasting it as a single optimization program. The resulting model is nonlinear, due to the bilinear terms $\OPprice{com}_{u,t} \OPexchangeOut_{u,t}$ and $\OPprice{com}_{u,t} \OPexchangeIn_{u,t}$ appearing in \eqref{upper:J_energy_component}. The recasting as a single optimization program relies on the fact that the lower level problem is a linear program, which can be replaced with its first-order necessary and sufficient Karush-Kuhn-Tucker conditions. Furthermore, the strong duality holds for a feasible linear program, and implies the complementary slackness \cite{bradley1977applied, boyd2004convexOptimization}. In order to avoid further nonlinearities due to complementary slackness conditions, the strong duality can thus be employed. Consequently, the bilevel model introduced in Section~\ref{sec:BilevelProgramming} can be recast as a single nonlinear optimization program composed of the following parts:
\begin{enumerate}
\item the objective function $\alpha$ of the upper level (to be maximized),
\item the upper-level constraints \eqref{upper:reserve_symmetric_def}-\eqref{upper:reserve_one_half_rule}, and $\alpha \geq 0$,
\item the lower-level constraints \eqref{lower:steerable_upper_bound}-\eqref{lower:reserve_symmetric_2},
\item the dual constraints of the lower level problem, reported in \ref{appendix:dual_contraints}, and
\item the strong duality condition for the lower level problem, described in \ref{appendix:strong_duality}.
\end{enumerate}
Notice that the lower level problem of the proposed bilevel model is always feasible, being feasible the decoupled solution where energy exchanges among the entities are set to zero. This is a consequence of the fact that the problems solved for each entity to compute the lower bounds $\profit{SU}_u$, are assumed to be feasible.

\section{Numerical results}\label{sec:NumericalResults}
Numerical results are divided in two categories. In Section~\ref{sec:Test1}, we report results on illustrative cases to demonstrate the proposed approach and the soundness of the solutions found. Then, in Section~\ref{sec:Test2}, we report a real test case to show the significant cost savings on a yearly scale for the community members, achieved by the proposed framework in a problem of real size. In all the tests, unless stated otherwise:
\begin{itemize}
\item $\initialCharge_{d} = \finalCharge_{d}=0$, $\forallb$;
\item $\maxExportToGrid=\infty$ and $\maxImportFromGrid=\infty$, $\forallu$;
\item entities can buy energy from, and sell energy to the grid at prices $\gridBuyPrice_t=0.15$~\euro/kWh and $\gridSalePrice_t=0.035$~\euro/kWh, respectively;
\item we check a posteriori that at each time step there is no simultaneous charging and discharging of any storage device.
\end{itemize}
All the results have been obtained using Pyomo 5.5 \cite{pyomoBook}, and Cplex 12.8 \cite{cplex2009v12}.

\subsection{Illustrative examples}\label{sec:Test1}
In the examples of this section, the duration of a time period is $\OPduration = 1$~h. Each entity is composed of only one device. For this reason, $\mathcal{D}^*_u = \{ u \}$, where the superscript~$*$ stands for the type of the device connected to entity~$u$ (\text{`nfl'}, \text{`nst'}, \text{`she'}, \text{`ste'}, or \text{`sto'}). In the frames of Figs.~\ref{fig:flows_case_1_p1}-\ref{fig:flows_case_5} showing energy flows and prices, the time index is removed from notation, because each frame refers to a single time period. In the tables of the same figures, the symbols in the first column should be intended with the subscript~$u$, when referred to entity~$u$. Similar to \eqref{upper:total_profit_u}, $\profit{SU,energy}_{u}$, $\profit{SU,peak}_{u}$ and $\profit{SU,reserve}_{u}$ denote the energy, peak and reserve component of $\profit{SU}_{u}$. On the other hand, the values in the second column, referred to the community, are the sums of the values referred to the entities along the same row.

\subsubsection{Excess of local generation, one time period} \label{sec:case1}
\begin{figure}[tb]
\centering
\begin{subfigure}{\linewidth}
	\centering
	\begin{boxedminipage}{\linewidth}
	\includegraphics[width=\linewidth]{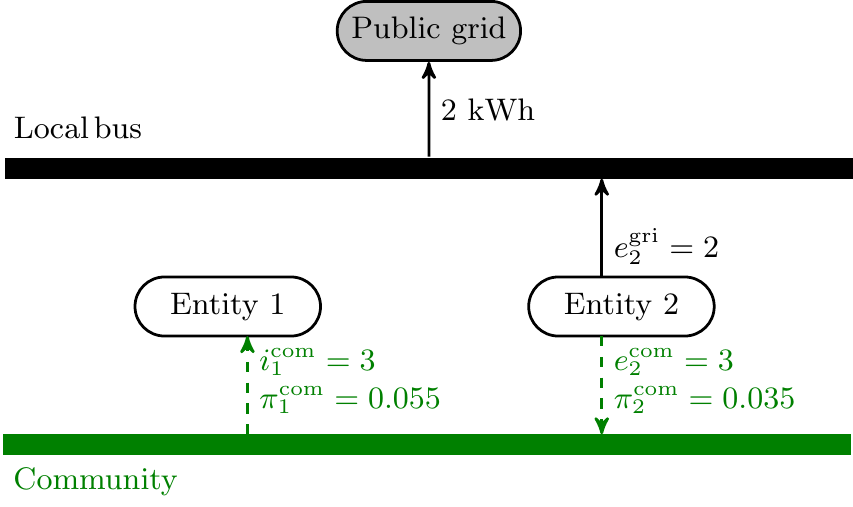}
	\end{boxedminipage}
	\caption{Energy flows and prices}
\end{subfigure}\\[1em]
\begin{subtable}{\linewidth}
	\centering
\begin{tabular}{|l|rrr|}
	\hline
	Entity & Com & 1 & 2 \\
	\hline
	\hline
	$\profit{}$        & 0.01                & -0.165              & 0.175               \\
	$\profit{SU}$        & -0.725              & -0.9                & 0.175               \\
	\hline
	$\profit{energy}$    & 0.01               & -0.165               & 0.175              \\
	$\profit{SU,energy}$    & -0.275               & -0.45                & 0.175              \\
	\hline
	$\profit{peak}$      & 0.0                 & 0.0                 & 0.0                 \\
	$\profit{SU,peak}$      & -0.45                & -0.45                & 0.0                 \\
	\hline
\end{tabular}
\caption{Summary of costs ($<0$) and revenues ($>0$)} \label{tab:case1}
\end{subtable}
\caption{Results of the example of Section~\ref{sec:case1}.}
\label{fig:flows_case_1_p1}
\end{figure}

The first example considers two entities, and encompasses only one time period ($t=1$). Entity~1 is a non-flexible load with $\nonFlexible_{1,1}=3$~kW, while entity~2 is a non-steerable generator with $\nonSteerable_{2,1}=5$~kW. Unitary peak power price and community operator fee are set to $\OPprice{peak} = 0.15$~\euro/kW and $\OPFee = 0.01$~\euro/kWh. Storage and reserve provision are not present in this example.

Figure~\ref{fig:flows_case_1_p1} shows the optimal solution, where entity~2 satisfies the demand of entity~1 through the community ($\OPexchangeOut_{2,1}=\OPexchangeIn_{1,1}=3$~kWh). The excess of generation is sold to the grid ($\exportGrid_{2,1}=2$~kWh) at price $\gridSalePrice_1=0.035$~\euro/kWh, which sets the community market price $\priceCom_{2,1}$ for entity~2. Indeed, the main grid represents the marginal unit, and its bid price defines the local market price for that entity. In this way, entity~2 does not improve its revenue as compared to acting individually, see Fig.~\ref{tab:case1}. Concerning entity~1, it buys energy from entity~2 at price $\priceCom_{1,1}=0.055$~\euro/kWh, which includes the market price $\priceCom_{2,1}$ of entity~2, and the fee $\OPFee$ due to the community microgrid operator, collected both on import $\OPexchangeIn_{1,1}$ and export $\OPexchangeOut_{2,1}$. This reflects the following relation, which is derived from \eqref{proof:slackness_1} and \eqref{proof:slackness_2} for an entity~$u$ with $\OPexchangeOut_{u,t}>0$, and an entity~$u'$ with $\OPexchangeIn_{u',t}>0$:
\begin{equation} \label{eq:twice_fee}
\priceCom_{u',t} = \priceCom_{u,t} + 2 \OPFee.
\end{equation}
From Fig.~\ref{tab:case1}, it is apparent that the reduction of the costs for entity~1 as compared to acting individually, is due to two reasons. First, it buys energy from entity~2 at a price which is lower than the selling price of the grid. Second, it avoids to pay the peak penalty, since the community is globally exporting energy to the grid.

\subsubsection{Shortage of local generation, one time period} \label{sec:case2}
\begin{figure}[tb]
	\centering
	\begin{subfigure}{\linewidth}
		\centering
	\begin{boxedminipage}{\linewidth}
		\includegraphics[width=\linewidth]{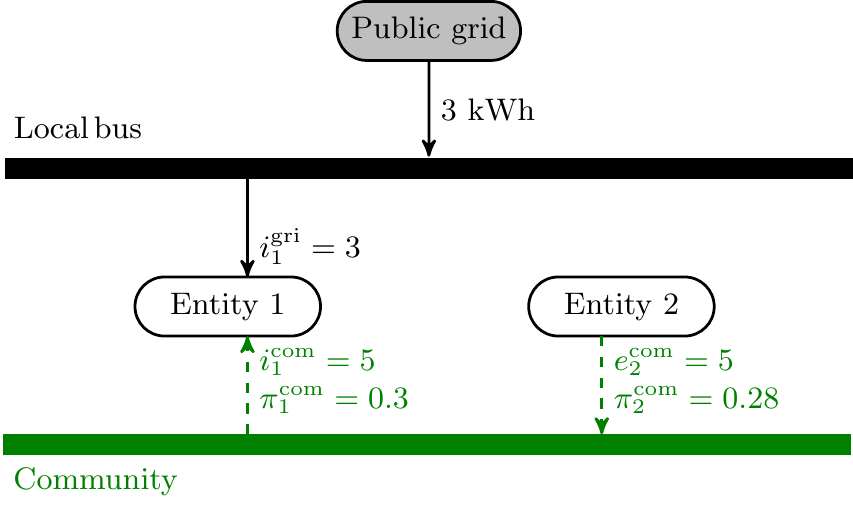}
	\end{boxedminipage}
		\caption{Energy flows and prices}
	\end{subfigure}\\[1em]
	\begin{subtable}{\linewidth}
		\centering
\begin{tabular}{|l|rrr|}
	\hline
	Entity            & Com                 & 1                   & 2                   \\
	\hline
	\hline
	$\profit{}$        & -1.0                & -1.95               & 0.95                \\
	$\profit{SU}$        & -2.225              & -2.4                & 0.175               \\
	\hline
	$\profit{energy}$    & -0.55                & -1.95                & 1.4                \\
	$\profit{SU,energy}$    & -1.025               & -1.2                 & 0.175              \\
	\hline
	$\profit{peak}$      & -0.45                & 0.0                 & -0.45                \\
	$\profit{SU,peak}$      & -1.2                 & -1.2                 & 0.0                 \\
	\hline
\end{tabular}
\caption{Summary of costs ($<0$) and revenues ($>0$)} \label{tab:case2}
\end{subtable}
\caption{Results of the example of Section~\ref{sec:case2}.}
\label{fig:flows_case_2_p1}
\end{figure}

The setup of this example is similar to that in Section~\ref{sec:case1}, with the only difference that entity~1 is now characterized by a consumption $\nonFlexible_{1,1}=8$~kW.

Figure~\ref{fig:flows_case_2_p1} shows the optimal solution. Entity~1 satisfies its demand buying energy from both entity~2 ($\OPexchangeOut_{2,1}=\OPexchangeIn_{1,1}=5$~kWh) and the grid ($\importGrid_{1,1}=3$~kWh). Since the grid supplies energy to entity~1, it acts as the marginal producer, setting the community market price $\priceCom_{1,1}$ for entity~1 as
\begin{equation} \label{eq:price_import_one_period}
\priceCom_{1,1} = \frac{\OPprice{peak}}{\Delta_T} + \gridBuyPrice_1.
\end{equation}
This can be derived from the complementary slackness conditions corresponding to the dual constraints~\eqref{dual:import_grid} and \eqref{appendix:pi_peak}, being $\importGrid_{1,1}>0$ and $\peak>0$. It is interesting to observe in Fig.~\ref{tab:case2} that the peak power cost of the community is totally assigned to entity~2, even though entity~2 is not importing energy from the grid. This is the effect of the optimization in the upper level, which, among all the feasible peak sharing policies, selects the one maximizing the minimum gain of the two entities.

\subsubsection{Storage, two time periods, no shared peak}\label{sec:case3}
\begin{figure}[tb]
	\centering
	\begin{subfigure}{\linewidth}
		\centering
		\begin{boxedminipage}{\linewidth}
		\includegraphics[width=\linewidth]{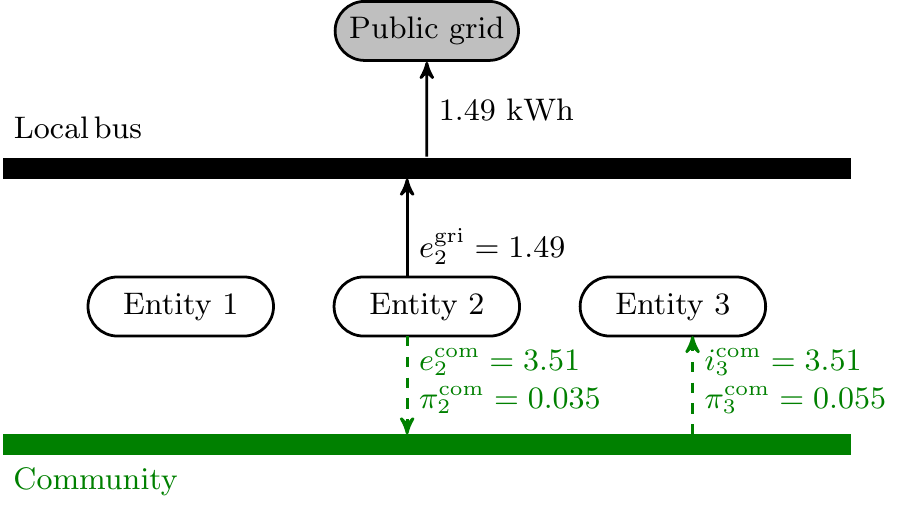}
		\end{boxedminipage}
		\caption{Energy flows and prices, time period 1}
	\end{subfigure}\\[1em]
	\begin{subfigure}{\linewidth}
		\centering
		\begin{boxedminipage}{\linewidth}
		\includegraphics[width=\linewidth]{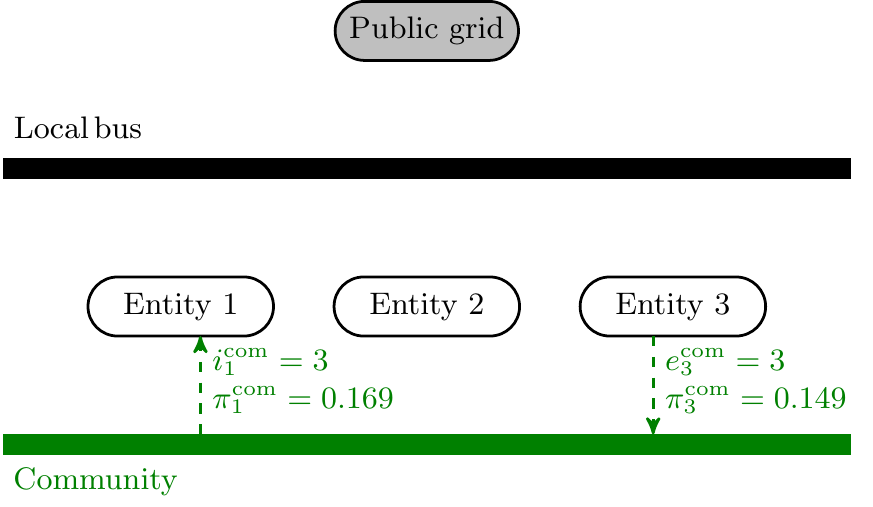}
		\end{boxedminipage}
		\caption{Energy flows and prices, time period 2}
	\end{subfigure}\\[1em]
	\begin{subtable}{\linewidth}
		\centering
		\begin{tabular}{|l|rrrr|}
			\hline
			Entity               & Com                 & 1                   & 2                   & 3                   \\
			\hline
            \hline
			$\profit{}$          & -0.331              & -0.506     & 0.175               & 0.0                 \\
			$\profit{SU}$        & -0.725              & -0.9                & 0.175               & 0.0                 \\
			\hline
			$\profit{energy  }$  & -0.331               & -0.506      & 0.175              & 0.0                 \\
			$\profit{SU,energy}$& -0.275               & -0.45                & 0.175              & 0.0                 \\
			\hline
			$\profit{peak}$      & 0.0                 & 0.0                 & 0.0                 & 0.0                 \\
			$\profit{SU,peak}$  & -0.45                & -0.45                & 0.0                 & 0.0                 \\
			\hline
		\end{tabular}
		\caption{Summary of costs ($<0$) and revenues ($>0$)}
		\label{tab:case3}
	\end{subtable}
	\caption{Results of the example of Section~\ref{sec:case3}.}
	\label{fig:flows_case_3}
\end{figure}

This example encompasses two time periods ($t=1,2$). Entity~1 is a non-flexible load with $\nonFlexible_{1,1}=0$~kW and $\nonFlexible_{1,2}=3$~kW. Entity~2 is a non-steerable generator with $\nonSteerable_{2,1}=5$~kW and $\nonSteerable_{2,2}=0$~kW. Different from previous examples, a third entity is added, characterized by a storage device with parameters $\maxcharge_3=12$~kWh, $\mincharge_3=0$~kWh, $\chargerate_3=\dischargerate_3=6$~kW, $\chargeEfficiency_3=0.9$ and $\dischargeEfficiency_3=0.95$. The unitary cost for battery usage is $\BSSsFee_{3} = 0.04$~\euro/kWh. All the other prices and fees remain unchanged with respect to Section~\ref{sec:case1}. Reserve provision is not present in this example.

Figure~\ref{fig:flows_case_3} shows the optimal solution. At time period $t=1$, entity~2 sells a part of its production to entity~3 ($\OPexchangeOut_{2,1}=\OPexchangeIn_{3,1}=3.51$~kWh), charging the storage up to a state of charge that is just enough so that entity~3 can fully satisfy the demand of entity~1 at time period $t=2$ ($\OPexchangeIn_{1,2}=\OPexchangeOut_{3,2}=3$~kWh), taking into account the round-trip efficiency of the storage ($\OPexchangeOut_{3,2}=\chargeEfficiency_3\dischargeEfficiency_3\OPexchangeIn_{3,1}$). Recall that $\initialCharge_{3} = \finalCharge_{3}=0$~kWh.
Concerning the market prices, at time period $t=1$ we are in a situation similar to that of Section~\ref{sec:case1}. The grid is the marginal unit, whose bid price defines the local market price for entity~2 ($\priceCom_{2,1} = \gridSalePrice_1=0.035$~\euro/kWh).
Then, the market price $\priceCom_{3,1}$ for entity~3 is determined by the relation~\eqref{eq:twice_fee} with $t=1$, $u=2$ and $u'=3$. On the other hand, at time period $t=2$, the market price $\priceCom_{3,2}$ for entity~3 is determined by replacing $d=3$ in the following relation:
\begin{equation} \label{eq:storage_prices}
\priceCom_{d,t+1} =\dfrac{\priceCom_{d,t}}{\chargeEfficiency_{d}\dischargeEfficiency_d} +2 \dfrac{\BSSsFee_{d}}{\dischargeEfficiency_d},
\end{equation}
which is obtained from the complementary slackness conditions corresponding to the dual constraints~\eqref{dual:charge}-\eqref{dual:sigmaZeta}, for a storage $d$ not providing reserve, charged at time period~$t$, and discharged at time period~$t+1$. Notice that \eqref{eq:storage_prices} is the minimum selling price that allows the storage unit to fully cover the costs for buying energy (including efficiency losses) and for its usage. Finally, the market price $\priceCom_{1,2}$ for entity~1 is determined by the relation~\eqref{eq:twice_fee} with $t=2$, $u=3$ and $u'=1$. The resulting price $\priceCom_{1,2}=0.169$~\euro/kWh at which entity~1 buys energy from the community, is greater than the price $\gridBuyPrice_2=0.15$~\euro/kWh at which it can buy energy from the grid. However, buying energy from the community avoids entity~1 to pay the penalty for the peak power (see Fig.~\ref{tab:case3}), with a significant cost saving.
\begin{remark}
If the maximum capacity of the storage device is reduced to ${\maxcharge_3=2}$~kWh, the storage is fully charged in period~1, namely $\OPSOC_{3,1}=2$~kWh. As a consequence, constraint~\eqref{lower:soc_upper_bound} is active, i.e., the storage is a scarce resource. Due to \eqref{dual:charge}-\eqref{dual:sigmaZeta}, and considering that $\phiSocMax \geq 0$, the additional term $\dfrac{\phiSocMax}{\dischargeEfficiency_d}$ appears in the right-hand side of \eqref{eq:storage_prices}. Thus, the storage gains 0.159~\euro~ by participating in the community. Notice, however, that the decrease of the storage capacity narrows the feasible set of the optimization problem, causing a reduction of the overall cost savings for the community.
\end{remark}

\subsubsection{Storage, two time periods, shared peak} \label{sec:case4}
\begin{figure}[tb]
	\centering
	\begin{subfigure}{\linewidth}
		\centering
	\begin{boxedminipage}{\linewidth}
		\includegraphics[width=\linewidth]{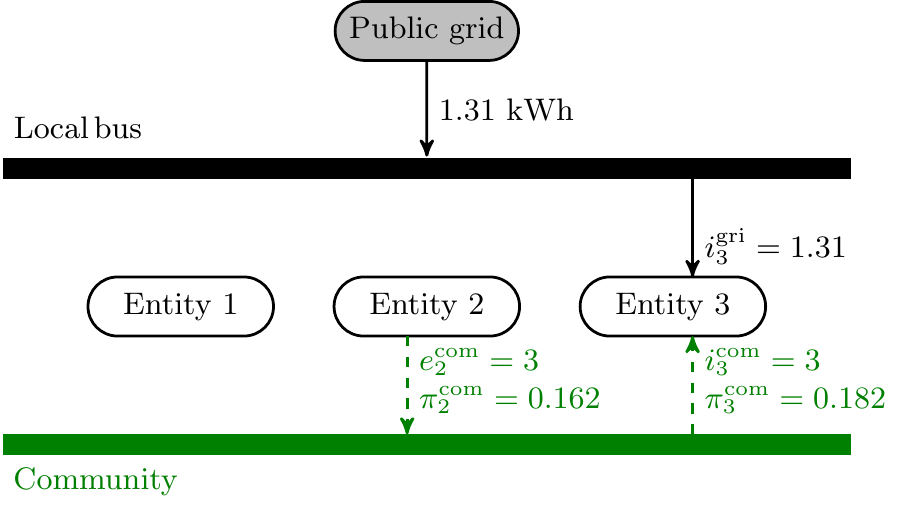}
	\end{boxedminipage}
		\caption{Energy flows and prices, time period 1}
	\end{subfigure}\\[1em]
	\begin{subfigure}{\linewidth}
		\centering
	\begin{boxedminipage}{\linewidth}
		\includegraphics[width=\linewidth]{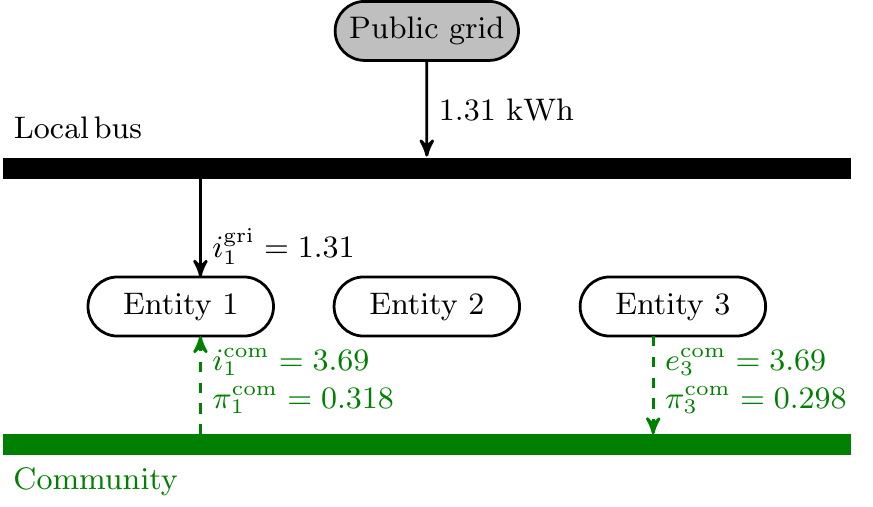}
	\end{boxedminipage}
		\caption{Energy flows and prices, time period 2}
	\end{subfigure}\\[1em]
	\begin{subtable}{\linewidth}
		\centering
		\begin{tabular}{|l|rrrr|}
			\hline
			Entity               & Com                 & 1                   & 2                   & 3                   \\
			\hline
			\hline
			$\profit{}$          & -1.101      & -1.63      & 0.487      & 0.0426     \\
			$\profit{SU}$        & -1.645      & -1.75               & 0.105               & 0.0                 \\
			\hline
			$\profit{energy  }$  & -0.838       & -1.368       & 0.487     & 0.0426    \\
			$\profit{SU,energy}$& -0.645       & -0.75                & 0.105              & 0.0                 \\
			\hline
			$\profit{peak}$      & -0.263       & -0.263      & 0.0                 & 0.0                 \\
			$\profit{SU,peak}$  & -1.0         & -1.0                 & 0.0                 & 0.0                 \\
			\hline
		\end{tabular}
		\caption{Summary of costs ($<0$) and revenues ($>0$)} \label{tab:case4}
	\end{subtable}
	\caption{Results of the example of Section~\ref{sec:case4}.}
	\label{fig:flows_case_4}
\end{figure}

The setup is similar to that of Section~\ref{sec:case3}, but in this case $\OPprice{peak} = 0.2$~\euro/kW, $\nonFlexible_{1,2}=5$~kW and $\nonSteerable_{2,1}=3$~kW.

Figure~\ref{fig:flows_case_4} shows the optimal solution. The production of entity~2 at time period $t=1$ is not enough to satisfy the demand of entity~1 at time period $t=2$. Hence, missing energy must be bought from the grid. In order to reduce the peak power cost, intuition suggests to buy the same amount of energy from the grid in both periods: in time period $t=1$ the energy imported from the grid is stored in the battery, while in time period $t=2$ it directly supplies entity~1. Writing down the energy balance under these constraints, it is straightforward to obtain
\begin{equation}
\importGrid_{3,1}=\importGrid_{1,2}=\frac{\nonFlexible_{1,2}-\chargeEfficiency_{3}\dischargeEfficiency_{3}\nonSteerable_{2,1}}{1+\chargeEfficiency_{3}\dischargeEfficiency_{3}}\simeq1.31~\text{kWh},
\end{equation}
which coincides with the solution found by the optimization model. To understand how the market prices are formed, notice that $\priceCom_{2,1}$ and $\priceCom_{3,1}$ satisfy \eqref{eq:twice_fee} for $u=2$, $u'=3$ and $t=1$, $\priceCom_{3,1}$ and $\priceCom_{3,2}$ satisfy \eqref{eq:storage_prices} for $d=3$ and $t=1$, and $\priceCom_{3,2}$ and $\priceCom_{1,2}$ satisfy \eqref{eq:twice_fee} for $u=3$, $u'=1$ and $t=2$. Moreover, by using \eqref{dual:import_grid} and \eqref{appendix:pi_peak}, it holds that:
\begin{equation} \label{eq:price_import_two_periods}
\priceCom_{3,1}+\priceCom_{1,2} = \frac{\OPprice{peak}}{\OPduration} + \gridBuyPrice_{1} + \gridBuyPrice_{2},
\end{equation}
which generalizes \eqref{eq:price_import_one_period} to the case of two entities importing energy from the grid, the one at time $t=1$, and the other at time $t=2$. We therefore have a system of four linear equations in the four unknowns $\priceCom_{2,1}$, $\priceCom_{3,1}$, $\priceCom_{1,2}$ and $\priceCom_{3,2}$, which admits the unique solution reported in Fig.~\ref{fig:flows_case_4}. It is apparent in Fig.~\ref{tab:case4} that all the three entities benefit from the market solution found at the community level.

\subsubsection{Flexible demand and generation, one time period} \label{sec:case6}
\begin{figure}[tb]
	\centering
	\begin{subfigure}{\linewidth}
		\centering
	\begin{boxedminipage}{\linewidth}
		\includegraphics[width=\linewidth]{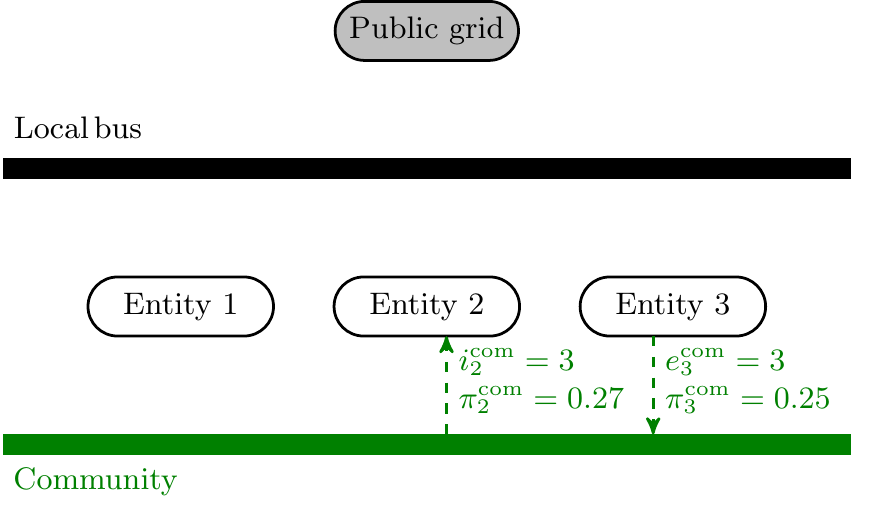}
	\end{boxedminipage}
		\caption{Energy flows and prices}
	\end{subfigure}\\[1em]
	\begin{subtable}{\linewidth}
		\centering
		\begin{tabular}{|l|rrrr|}
			\hline
			Entity            & Com                 & 1                   & 2                   & 3                   \\
			\hline
			\hline
			$\profit{}$        & -1.31               & -0.5                & -0.81               & 0.0                 \\
			$\profit{SU}$        & -1.4                & -0.5                & -0.9                & 0.0                 \\
			\hline
			$\profit{energy}$    & -1.31                & -0.5                 & -0.81                & 0.0                 \\
			$\profit{SU,energy}$    & -0.95                & -0.5                 & -0.45                & 0.0                 \\
			\hline
			$\profit{peak}$      & 0.0                 & 0.0                 & 0.0                 & 0.0                 \\
			$\profit{SU,peak}$      & -0.45                & 0.0                 & -0.45                & 0.0                 \\
			\hline
		\end{tabular}
		\caption{Summary of costs ($<0$) and revenues ($>0$)} \label{tab:case6}
	\end{subtable}
	\caption{Results of the example of Section~\ref{sec:case6}.}
	\label{fig:flows_case_6}
\end{figure}

This example encompasses only one time period ($t=1$). Entities~1 and 2 are sheddable loads with $\sheddable_{1,1}=5$~kW and $\sheddable_{2,1}=3$~kW, respectively. The corresponding load shedding costs amount to $\sheddingPrice_{1,1}=0.1$ and $\sheddingPrice_{2,1}=0.4$~\euro/kWh. Entity~3 is a steerable generator with $\steerable_{3,1}=4$~kW, and generation cost $\steerablePrice_{3,1}=0.25$~\euro/kWh. All the other prices and fees remain unchanged with respect to Section~\ref{sec:case1}. Storage and reserve are not present in this example.

Figure~\ref{fig:flows_case_6} shows the optimal solution. When entities act individually, entity~1 sheds the whole demand, because its cost of load shedding is less than the cost of buying energy from the grid, including the peak power cost, namely
\begin{equation} \label{eq:cost_buy_energy_from_grid}
\frac{\OPprice{peak}}{\Delta_T} + \gridBuyPrice_1 = 0.3~\text{\euro/kWh.}
\end{equation}
On the other hand, entity~2 consumes its maximum power, because its cost of load shedding is greater than \eqref{eq:cost_buy_energy_from_grid}. Entity~3 does not produce, because its generation cost is not compensated by the price at which it may sell energy to the grid, i.e., $\gridSalePrice_1 = 0.035$~\euro/kWh. When the entities are within the community, the best option for entity~1 is still to shed the whole demand, while entity~2 is fully supplied by entity~3 through the community. In this way, peak power costs are avoided. The market price $\priceCom_{3,1}$ for entity~3 (which is the marginal unit in this case) coincides with its generation cost $\steerablePrice_{3,1}$, i.e., the minimum price at which entity~3 does not incur in losses. The market price for entity~2 is determined by \eqref{eq:twice_fee} for $u=3$, $u'=2$ and $t=1$. It can be observed in Fig.~\ref{tab:case6} that the gain of the community entirely corresponds to the gain of entity~2.

\subsubsection{Steerable generation, one time period, reserve} \label{sec:case5}
\begin{figure}[tb]
	\centering
	\begin{subfigure}{\linewidth}
		\centering
	\begin{boxedminipage}{\linewidth}
		\includegraphics[width=\linewidth]{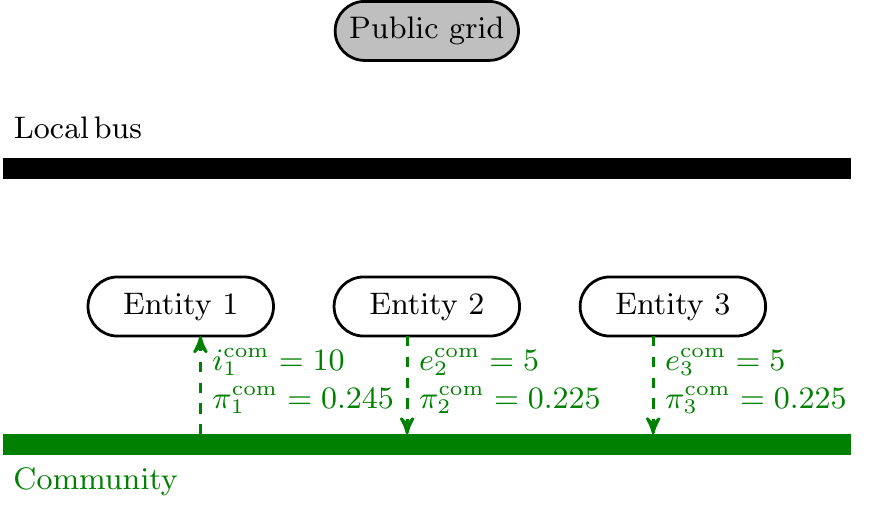}
	\end{boxedminipage}
		\caption{Energy flows and prices}
	\end{subfigure}\\[1em]
	\begin{subtable}{\linewidth}
		\centering
\begin{tabular}{|l|rrrr|}
	\hline
	Entity              & Com                 & 1                   & 2                   & 3                   \\
	\hline
	\hline
	$\profit{}$         & 0.575               & -2.45               & 1.303               & 1.722                 \\
	$\profit{SU}$       & -1.4125             & -3.0                & 0.5375              & 1.05                \\
	\hline
	$\profit{energy}$   & -0.425               & -2.45                & 1.025              & 1.0                \\
	$\profit{SU,energy}$    & -1.4125              & -1.5                 & 0.0375             & 0.05               \\
	\hline
	$\profit{peak}$     & 0.0                 & 0.0                 & 0.0                 & 0.0                 \\
	$\profit{SU,peak}$ & -1.5                 & -1.5                 & 0.0                 & 0.0                 \\
	\hline
	$\profit{reserve }$ & 1.0                 & 0.0                 & 0.278                 & 0.722                 \\
	$\profit{SU,reserve}$ & 1.5                 & 0.0                 & 0.5                 & 1.0                 \\
	\hline
\end{tabular}
		\caption{Summary of costs ($<0$) and revenues ($>0$)} \label{tab:case5}
	\end{subtable}
	\caption{Results of the example of Section~\ref{sec:case5}.}
	\label{fig:flows_case_5}
\end{figure}

This example encompasses only one time period ($t=1$). Entity~1 is a non-flexible load with $\nonFlexible_{1,1}=10$~kW. Entities~2 and 3 are steerable generators with $\steerable_{2,1}=5$~kW and $\steerable_{3,1}=10$~kW, respectively. Both generators have relatively low generation costs, namely $\steerablePrice_{2,1}=0.02$ and $\steerablePrice_{3,1}=0.025$~\euro/kWh. Reserve price is set to $\reservePrice = 0.2$~\euro/kW. All the other prices and fees remain unchanged with respect to Section~\ref{sec:case1}. Storage is not present in this example.

Figure~\ref{fig:flows_case_5} shows the optimal solution. Given the high reserve price and the low price at which energy is sold to the grid, when they act individually, entities~2 and 3 produce at half of their capacity, thus maximizing the revenue from symmetric reserve provision. On the other hand, when they are within the community, suitable pricing of energy exchanges makes selling energy to the grid no more convenient for entities~2 and 3. According to the marginal pricing framework, for both entities the market price would be the cost of providing one additional unit of energy. In this case, it is given by the sum of the corresponding generation cost, and the missing profit from providing reserve (i.e., the opportunity cost). Moreover, for two entities both exporting energy to the community, the market prices are equal, as follows from the complementary slackness conditions corresponding to the dual constraints~\eqref{dual:export_com}. Since $\steerablePrice_{3,1} > \steerablePrice_{2,1}$, entity~3 is the marginal unit, and defines the price. Indeed, the optimization returns
\begin{equation}
\priceCom_{2,1} = \priceCom_{3,1} = \frac{\reservePrice}{\OPduration} + \steerablePrice_{3,1} = 0.225~\text{\euro/kWh.}
\end{equation}
At this price, it is convenient for entity~2 to steer all its available power, while entity~3 produces at half of its capacity, still guaranteeing its maximum amount of symmetric reserve. The market price for entity~1 is determined by \eqref{eq:twice_fee} for $u=2$ (or $u=3$), $u'=1$ and $t=1$. It can be observed in Fig.~\ref{tab:case5} that all the three entities benefit from the market solution found at the community level. Notice that the community assigns a portion of the revenue from reserve provision also to entity~2, although entity~2 does not actually contribute to symmetric reserve (it produces at its maximum capacity).

\subsection{The MeryGrid project test case}\label{sec:Test2}
\begin{figure}[!t]
\begin{subfigure}{\linewidth}
\centering
\includegraphics[width=1\linewidth]{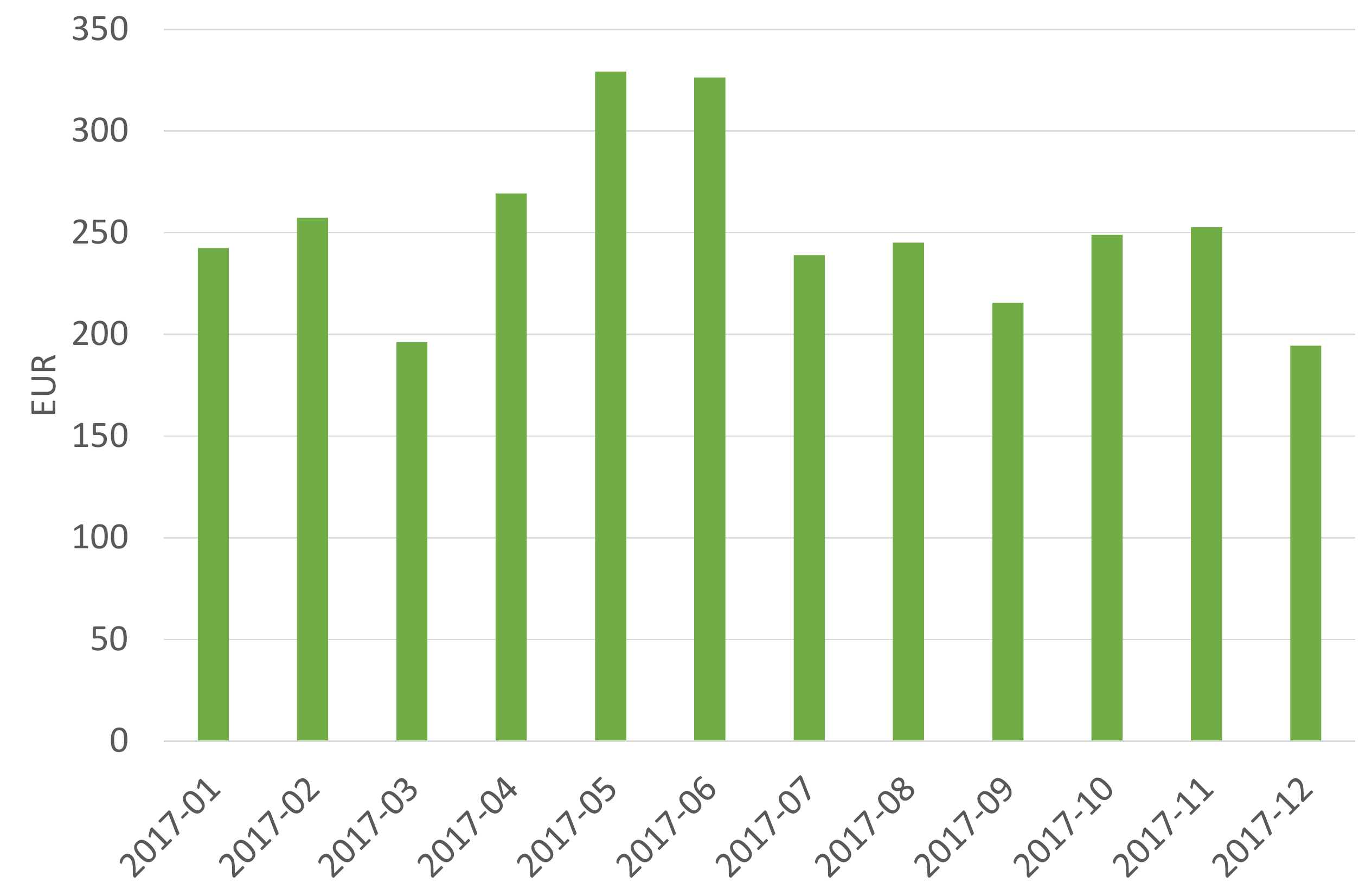}
\caption{Monthly cost of using storage.}
\label{fig:storage_fee_NL}
\end{subfigure}\\
\begin{subfigure}{\linewidth}
\centering
\includegraphics[width=1\linewidth]{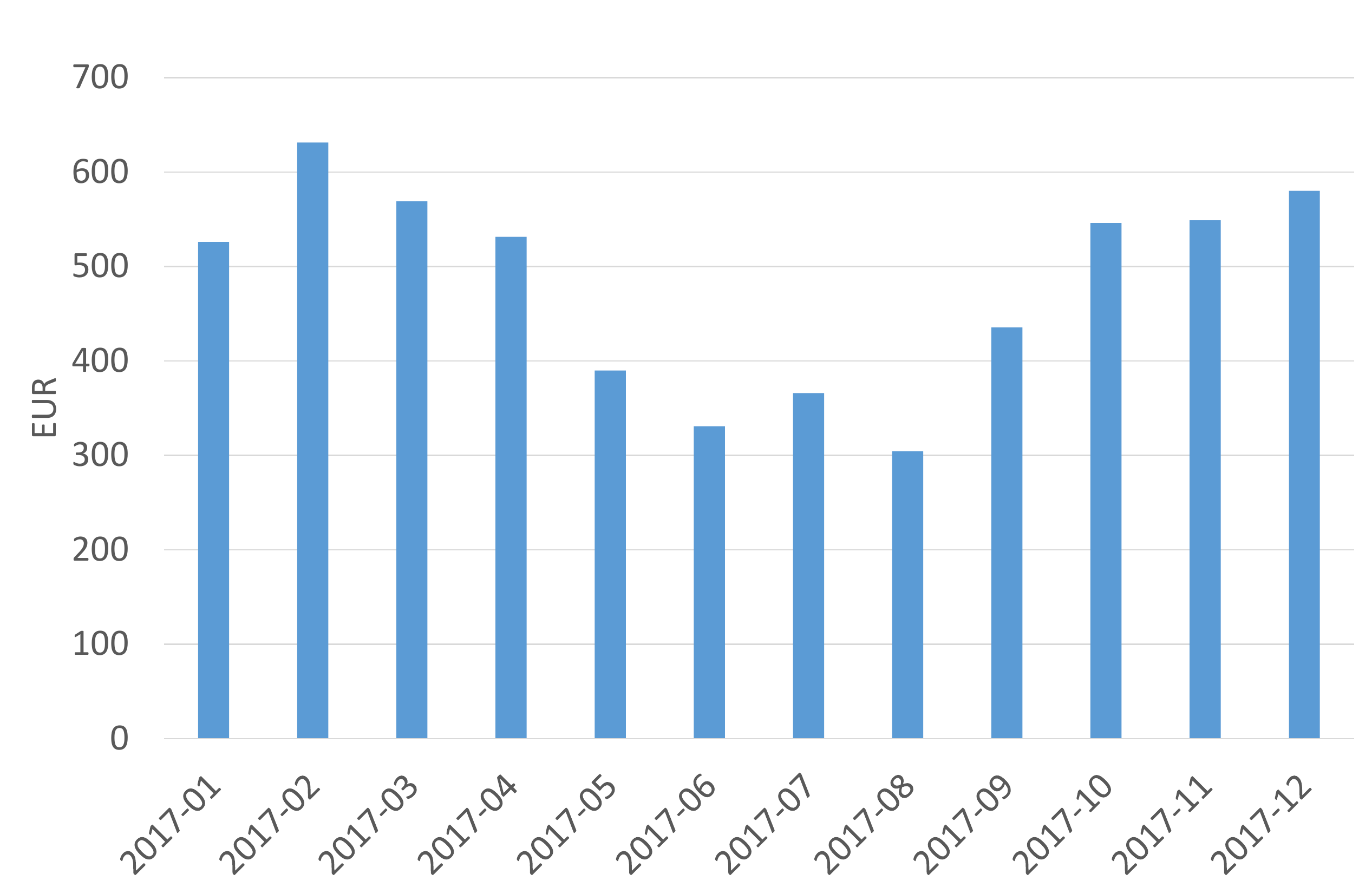}
\caption{Monthly fee paid to community operator.}
\label{fig:operator_fee_NL}
\end{subfigure}
\caption{Costs for the community.}
\label{fig:real_costs}
\end{figure}

This test case is inspired by the MeryGrid project \cite{cornelusse2017efficient}, which is a real pilot project currently under implementation in the Walloon region of Belgium, in agreement with the energy regulator, the DSO, the companies (entities) that constitute the community, and their energy retailers.

There are four entities. Entity~1 is a non-flexible load. Entity~2 is composed of a non-flexible load and a (non-steerable) photovoltaic plant with 70~kWp of installed power. Entity~3 is composed of a non-flexible load and a hydro-electric plant with a capacity of about 200~kVA, treated as a non-steerable generator. Finally, entity~4 is a 270~kWh battery storage system, with 150~kW and 300~kW of maximum charging and discharging power, respectively, and charging and discharging efficiencies both equal to 0.95.
\begin{table}[!b]
\centering
\setlength{\tabcolsep}{5pt}
\setlength\extrarowheight{2pt}
\caption{Statistical indexes of load and generation profiles (in kW).} \label{tab:stat_profiles}
\begin{tabular}{|c|c|c|c|c|c|}
\hline
Entity & Device & Max & Min & Avg & Std \\ \hline \hline
1 & \text{Load} & 288.40 & 0.00 & 23.13 & 29.84 \\ \hline
\multirow{2}{*}{2} & \text{Load} & 91.20 & 0.00 & 16.76 & 17.94  \\
& \text{Gen} & 68.96 & 0.00 & 3.66 & 9.63 \\ \hline
\multirow{2}{*}{3} & \text{Load} & 271.64 & 0.00 & 9.02 & 24.86  \\
& \text{Gen} & 189.82 & 0.00 & 74.64 & 51.86 \\ \hline
\end{tabular}
\end{table}

Demand and generation profiles of the three loads and two generators are available for the whole year~2017 with time step $\OPduration=15$~min. The maximum, minimum and average values of these profiles (in kW), as well as their standard deviation, over the considered year are reported in Table~\ref{tab:stat_profiles}. One instance of the proposed optimization model is solved for every day of the year. For each instance, the number of time periods is $T=96$, while the number of optimization variables and constraints is around 5500 and 5900, respectively. The average computation time for a single instance is around 6.36~s on a 4-core 3.6~GHz Intel i7-7700 CPU with 32~GB of RAM. Unitary peak power price, community operator fee, and cost of storage usage are those defined in Sections~\ref{sec:case1} and \ref{sec:case3}. The prices of energy imported from and exported to the grid are assumed to be constant over the whole period, with values given at the beginning of Section~\ref{sec:NumericalResults}. For the storage system, the state of charge at the beginning and at the end of the day are set equal to half of its maximum capacity, namely 135~kWh. Reserve provision is ignored. Indeed, only the storage system is able to provide symmetric reserve in this example, and when increasing the price of reserve, the storage system tends to provide only reserve, and not to participate in the community.

\begin{figure}[!t]
\begin{subfigure}{\linewidth}
\centering
\includegraphics[width=1\linewidth]{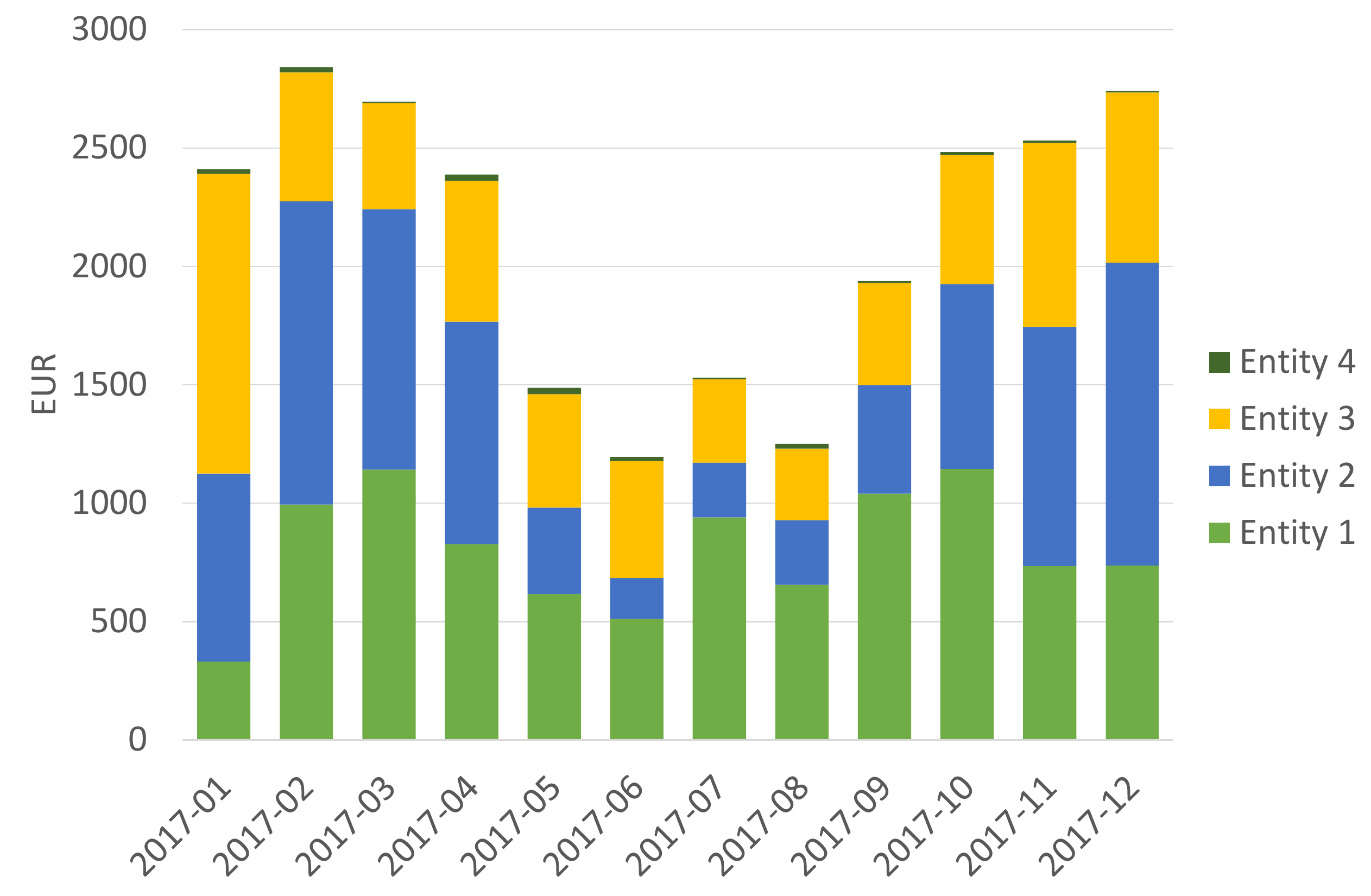}
\caption{Monthly gains.}
\label{fig:summary_bar_NL}
\end{subfigure}\\
\begin{subfigure}{\linewidth}
	\centering
	\includegraphics[width=\linewidth]{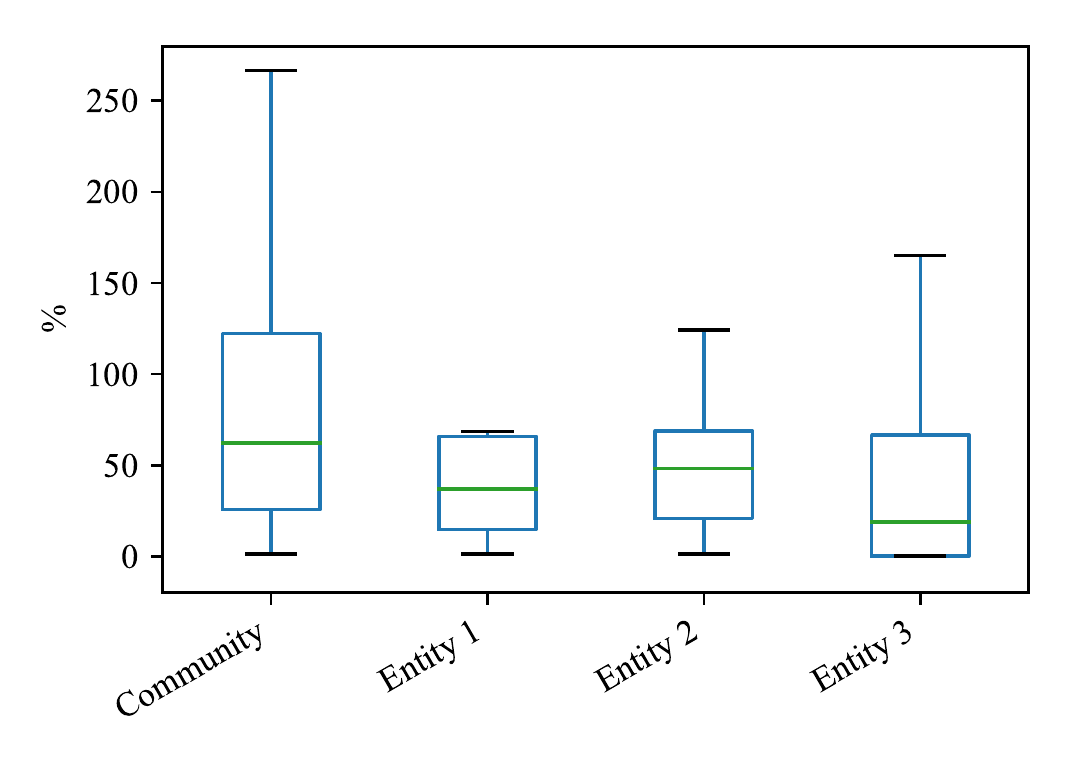}
	\caption{Boxplots of percent daily gains.}
	\label{fig:summary_box_NL}
\end{subfigure}
\caption{Gains for the community and for each entity.}
\label{fig:real_gains}
\end{figure}

Results for this test case are summarized in Figs.~\ref{fig:real_costs} and \ref{fig:real_gains}. Figure~\ref{fig:storage_fee_NL} shows the monthly cost of using the battery storage, while Fig.~\ref{fig:operator_fee_NL} shows the fee paid monthly to the community operator. Since unitary cost of storage usage and community operator fee are constant over the whole year, these costs are directly proportional to the amounts of energy exchanged monthly with the storage unit, and among the entities of the community, respectively. Figure~\ref{fig:summary_bar_NL} shows the stacked bar plot of the monthly gains of all the entities. The height of each bar represents the monthly gain of the community as a whole. Notice that, for each entity~$u$, the monthly (resp. yearly) gain is the cumulative value at the end of a month (resp. year) of the daily gains $\profit{}_u-\profit{SU}_u$. All the entities except entity~4 have nonnegligible gains. On a yearly scale, entities~1 and 2 enjoy cost savings amounting to 28.36\% and 37.98\%, respectively, while entity~3 has a revenue increase of 73.47\%. On the other hand, entity~4 gains only 180~\euro~at the end of the year. This can be explained in analogy with the toy example of Section~\ref{sec:case3}. Entity~4 (connected to the storage unit) typically sells energy to the community at a price which only balances the costs for buying energy (including efficiency losses) and for its usage. Finally, boxplots of percent daily gains are shown in Fig.~\ref{fig:summary_box_NL} for the community and for entities~1, 2 and 3. These percent gains vary significantly from day to day, but the median value is between 19\% and 62\% in all the cases reported. For entity~4 percent daily gains cannot be computed, because its daily profit when acting individually is always zero (recall that $\gridBuyPrice_{t} > \gridSalePrice_{t}$).

\section{Conclusions}\label{sec:Conclusions}
This paper has introduced a modelling framework, based on bilevel programming, for structuring and solving the internal local market of a community microgrid, composed of entities which may exchange energy and services among themselves. The proposed framework has two main features. First, it guarantees a market-oriented pricing of energy exchanges within the community by implementing a social welfare maximization approach based on the marginal pricing scheme. Second, by imposing a Pareto-superior condition, it ensures that no entity is penalized by participating in the community, as compared to acting individually, which is a fundamental requirement for building a solid and long-lasting community. Numerical results obtained on a real test case implemented in Belgium show that the proposed framework is effective in enhancing the welfare of the community members, which gain between 28\% and 74\% on a yearly scale, with respect to the case when they act individually.

The approach of this paper can be readily extended to handle curtailment of power generation, different sharing schemes of both community reserve and community peak power, and more advanced demand side models, as long as the lower level problem of the proposed bilevel model remains a linear program.

Further work aims at studying advanced formulations of the upper level problem, in order to achieve a better sharing of the benefits of the community among all the entities, as well as analyzing possible strategic behaviors of the participants. Current research is also focusing on the introduction of smart buildings, electric vehicles, and community-of-communities concepts in the proposed framework. Finally, another frontier is to look for integration of communities into existing electricity markets \cite{Zepter2019}.

\section*{Acknowledgement}
Bertrand Corn\'elusse thanks Nethys, Li\`{e}ge, Belgium, for its support.

\appendix

\section{Lower level dual constraints}\label{appendix:dual_contraints}

This appendix reports the dual constraints of the lower level problem described in Section~\ref{sec:LowerLevel}. These constraints are obtained by applying standard tools of duality theory in linear programming.

The dual constraint of $\steer_{d,t} \geq 0$ is defined as:
\begin{align}
&\phiSteerMax-\priceCom_{u,t}\OPduration\steerable_{d,t} +\steerable_{d,t}(\rhoReserveSymInc-\rhoReserveSymDec) \geq - \OPprice{ste}_{d,t} \steerable_{d,t}\OPduration
\end{align}
The dual constraint of $\shed_{d,t} \geq 0$ is defined as:
\begin{align}
&\phiShedMax-\priceCom_{u,t}\OPduration\sheddable_{d,t} +\sheddable_{d,t}(\rhoReserveSymInc-\rhoReserveSymDec) \geq - \OPprice{she}_{d,t} \sheddable_{d,t} \OPduration
\end{align}
The dual constraints of $\charge_{d,t}  \geq 0$ and $\discharge_{d,t}  \geq 0$ are respectively:
\begin{align}
&\phiChargeMax -\OPduration\chargerate_{d}\chargeEfficiency_{d}\sigma_{d,t} +\OPduration\chargerate_{d}\priceCom_{u,t} + \notag \\
 & \hspace*{100pt} +\chargerate_d \phiReserveBssMaxDec \geq - \BSSsFee_{d} \OPduration \chargerate_{d} \chargeEfficiency_{d} \label{dual:charge}\\
&\phiDischargeMax  +\dfrac{\OPduration\dischargerate_{d}}{\dischargeEfficiency_{d}}\sigma_{d,t}  -\OPduration\dischargerate_{d}\priceCom_{u,t} + \notag \\
 & \hspace*{100pt} + \dischargerate_d \phiReserveBssMaxInc \geq -\BSSsFee_{d} \dfrac{\OPduration \dischargerate_{d}}{\dischargeEfficiency_d}
\end{align}
The dual constraints of $\OPSOC_{d,t} \in \mathbb{R}$ for $t \in \{1,\ldots,T-1\}$ and $t = T$ are respectively:
\begin{align}
&\phiSocMax-\phiSocMin+\sigma_{d,t}-\sigma_{d,t+1} - \dfrac{\kappaReserveBssSocInc\dischargeEfficiency_{d}}{\OPduration} + \dfrac{\kappaReserveBssSocDec}{\chargeEfficiency_{d}\OPduration} = 0\\
&\phiSocMaxT-\phiSocMinT+\sigma_{d,T} +\zeta_d - \dfrac{\kappaReserveBssSocIncT\dischargeEfficiency_{d}}{\OPduration} + \dfrac{\kappaReserveBssSocDecT}{\chargeEfficiency_{d}\OPduration} = 0 \label{dual:sigmaZeta}
\end{align}
The dual constraints of $\OPexchangeOut_{u,t} \geq 0$ and $\OPexchangeIn_{u,t} \geq 0$ are respectively:
\begin{align}
\priceCom_{u,t}-\mu_{t}  &\geq -\OPFee \label{dual:export_com}\\
-\priceCom_{u,t}+\mu_{t} &\geq -\OPFee \label{dual:import_com} &
\end{align}
The dual constraints of $\exportGrid_{u,t} \geq 0$ and $\importGrid_{u,t} \geq 0$ are respectively:
\begin{align}
&\priceCom_{u,t}-\phiPeak/\OPduration + (\phiGridCapExport-\phiGridCapImport)/\OPduration \geq\gridSalePrice_{t} \label{dual:export_grid}\\
&-\priceCom_{u,t}+\phiPeak/\OPduration + (\phiGridCapImport - \phiGridCapExport)/\OPduration \geq-\gridBuyPrice_{t} \label{dual:import_grid}
\end{align}
The dual constraints of $\reserveBSSInc  \geq 0 $ and $\reserveBSSDec  \geq 0$ are respectively:
\begin{align}
&\kappaReserveBssSocInc+\phiReserveBssMaxInc-\rhoReserveSymInc\geq0\\
&\kappaReserveBssSocDec+\phiReserveBssMaxDec-\rhoReserveSymDec\geq0
\end{align}
The dual constraint of $\OPreserve{sym} \geq 0$ is defined as:
\begin{align}
&\sum_{t \in \OPperiods} (\rhoReserveSymInc+\rhoReserveSymDec)\geq\OPprice{res}
\end{align}
The dual constraint of $\peak \geq 0$ is defined as:
\begin{align} \label{appendix:pi_peak}
&-\sum_{t \in \OPperiods}\phiPeak\geq-\OPprice{peak}
\end{align}

\section{Strong duality}\label{appendix:strong_duality}
The following expression represents the objective function of the lower level dual problem:
\begin{align}
&\sum_{t \in \OPperiods} \bigg(\sum_{d\in \sheddableDevices} \phiShedMax + \sum_{d\in \steerableDevices} \phiSteerMax + \sum_{d\in\BSSs} \phiChargeMax +\sum_{d\in\BSSs} \phiDischargeMax\bigg)\notag\\
&+ \sum_{t \in \OPperiods} \sum_{d\in\BSSs} \left(\phiSocMax\maxcharge_{d}-\phiSocMin\mincharge_{d}\right)\notag\\
&-\sum_{t \in \OPperiods} \sum_{u \in \users} \priceCom_{u,t} \bigg(\sum_{d\in \nonsteerableDevices_u} \OPduration\nonSteerable_{d,t}  - \sum_{d\in \sheddableDevices_u}\OPduration\sheddable_{d,t} + \notag \\ & \qquad - \sum_{d\in\nonflexibleDevices_u}\OPduration\nonFlexible_{d,t}\bigg)+\sum_{d\in\BSSs}\sigma_{d,1}\initialCharge_{d} + \sum_{d\in\BSSs}\zeta_d\finalCharge_{d}\notag\\
&+\sum_{t \in \OPperiods}\sum_{d\in\BSSs}\bigg(-\kappaReserveBssSocInc\dischargeEfficiency_{d}\mincharge_{d}/\OPduration + \phiReserveBssMaxInc \dischargerate_d\notag\\
&\qquad+\kappaReserveBssSocDec\maxcharge_{d}/(\chargeEfficiency_{d}\OPduration) + \phiReserveBssMaxDec \chargerate_d\bigg)\notag\\
&+\sum_{t \in \OPperiods}\rhoReserveSymInc\bigg(\sum_{d\in \steerableDevices}\steerable_{d,t} + \sum_{d\in \sheddableDevices} \sheddable_{d,t} \bigg) \notag\\
&+\sum_{u \in \users} \sum_{t \in \OPperiods} (\phiGridCapExport\maxExportToGrid + \phiGridCapImport\maxImportFromGrid) \label{lhs_strong_duality}
\end{align}
The strong duality condition requires the identity, at the optimum, of the values of the primal objective function \eqref{lower:objective_function} and the dual objective function \eqref{lhs_strong_duality}.

\section{Proofs}
\subsection{Sum of bilinear terms involving community prices and community import/export}\label{appendix:proof_sum_bilinear_terms}
This appendix shows that identity~\eqref{lower:cost_identity} holds at the optimum of the lower level problem. First, we notice that
\begin{equation} \label{proof:slackness_1}
\OPexchangeOut_{u,t} \OPprice{com}_{u,t} = \OPexchangeOut_{u,t} (\mu_{t} - \OPFee).
\end{equation}
If $\OPexchangeOut_{u,t}=0$, the identity is trivial. If $\OPexchangeOut_{u,t}>0$, it follows from the complementary slackness condition corresponding to the dual constraint~\eqref{dual:export_com}, implying $\priceCom_{u,t} = \mu_{t} -\OPFee$. Similarly,
\begin{equation} \label{proof:slackness_2}
\OPexchangeIn_{u,t} \OPprice{com}_{u,t} = \OPexchangeIn_{u,t} (\mu_{t} + \OPFee),
\end{equation}
where the identity is trivial when $\OPexchangeIn_{u,t}=0$, while it follows from the complementary slackness condition corresponding to the dual constraint~\eqref{dual:import_com}, implying $\priceCom_{u,t} = \mu_{t} + \OPFee$, when $\OPexchangeIn_{u,t}>0$. By using \eqref{proof:slackness_1} and \eqref{proof:slackness_2}, we obtain:
\begin{align*}
&\sum_{t \in \OPperiods} \sum_{u\in\users} \OPprice{com}_{u,t} \left( \OPexchangeOut_{u,t} - \OPexchangeIn_{u,t} \right) = \\
&=\sum_{t \in \OPperiods} \sum_{u\in\users} \OPexchangeOut_{u,t} (\mu_{t} - \OPFee) - \sum_{t \in \OPperiods} \sum_{u\in\users} \OPexchangeIn_{u,t} (\mu_{t} + \OPFee)=\\
&=\sum_{t \in \OPperiods} \sum_{u\in\users} \mu_{t} (\OPexchangeOut_{u,t} - \OPexchangeIn_{u,t})
- \OPFee \sum_u \sum_t  (\OPexchangeOut_{u,t}
+ \OPexchangeIn_{u,t})\\
&=- \OPFee \sum_{t \in \OPperiods} \sum_{u\in\users}  \left(\OPexchangeOut_{u,t}
+ \OPexchangeIn_{u,t}\right),
\end{align*}
where the last equality holds due to \eqref{lower:export_import_balance}.

\subsection{No simultaneous import/export}\label{appendix:proof_simultanelus_import_export}
This section shows that no simultaneous import from or export to the grid is possible by the same entity at the same time, as long as $\gridSalePrice_t \neq \gridBuyPrice_t$.

Indeed, from the dual constraints \eqref{dual:export_grid} and \eqref{dual:import_grid}, due to the complementary slackness conditions, the following hold:
\begin{align}
&\text{if }  \exportGrid_{u,t} > 0  &&\text{then} &&  \priceCom_{u,t} - \dfrac{\phiPeak}{\OPduration} + \dfrac{\phiGridCapExport-\phiGridCapImport}{\OPduration} = \gridSalePrice_{t}  \label{proof:slackness_exp_grid}\\
&\text{if }  \importGrid_{u,t} > 0    &&\text{then} && \priceCom_{u,t} - \dfrac{\phiPeak}{\OPduration} + \dfrac{ \phiGridCapExport - \phiGridCapImport }{\OPduration} = \gridBuyPrice_{t}. \label{proof:slackness_imp_grid}
\end{align}
Then, if $\gridSalePrice_t \neq \gridBuyPrice_t$, the conditions \eqref{proof:slackness_exp_grid} and \eqref{proof:slackness_imp_grid} cannot hold simultaneously. Therefore, the variables $\exportGrid_{u,t}$ and  $\importGrid_{u,t}$ cannot be both strictly positive at the same time.

Notice that, for similar considerations on the conditions \eqref{proof:slackness_1} and \eqref{proof:slackness_2}, as long as $\OPFee \neq 0$, it is not even possible to have simultaneous import from and export to the community by the same entity.

%
\bibliographystyle{elsarticle-num}
\bibliography{mybib}

\end{document}